\documentclass[aps,prd,a4paper]{revtex4}
\usepackage{ulem}
\usepackage{color}
\usepackage{graphicx}
\usepackage{epsfig}
\usepackage{graphicx,epsfig,amsmath,bm,array,subfigure,amssymb}
\usepackage{slashed}
\usepackage{hyperref}

\newcommand{\be}{\begin{equation}}
\newcommand{\ee}{\end{equation}}
\newcommand{\bea}{\begin{eqnarray}}
\newcommand{\eea}{\end{eqnarray}}
\def\be{\begin{equation}}
\def\ee{\end{equation}}
\def\bearr{\begin{eqnarray}}
\def\eearr{\end{eqnarray}}
\def\zbf#1{{\bf {#1}}}
\def\bfm#1{\mbox{\boldmath $#1$}}

\def\vec#1{\mathchoice
{\mbox{\boldmath $#1$}}
{\mbox{\boldmath $#1$}}
{\mbox{\boldmath $\scriptstyle #1$}}
{\mbox{\boldmath $\scriptscriptstyle #1$}}
}

\newcommand{\PRD}[3]{Phys.\ Rev.\ D\ {\bf #1},\ #2 (#3)}

\begin{document}

\title{Heavy quark diffusion in a Polyakov loop plasma}
\author{Balbeer Singh$^{1,2}$, Aman Abhishek$^{1,2}$,  Santosh K. Das$^3$ , Hiranmaya Mishra$^1$}
\affiliation{$^{1}$Theory Division, Physical Research Laboratory, 
Navrangpura, Ahmedabad 380 009, India}
\affiliation{$^2$ Indian Institute of Technology Gandhinagar
Gandhinagar 382 355, Gujarat, India}
\affiliation{$^{3}$ School of Physical Science, Indian Institute of Technology Goa, Ponda-403401, Goa,  India}
\date{\today}
\begin{abstract}
We calculate the transport coefficients, drag and momentum diffusion,  of a heavy quark in a thermalized plasma of light quarks in the background of Polyakov loop. Quark thermal mass and the gluon Debye mass are calculated in a non-trivial Polyakov loop background. The constituent quark masses and the Polyakov loop is estimated within a Polyakov loop quark meson (PQM) model. The relavant scattering amplitudes for heavy quark and light partons in the background of Polyakov loop has been estimated within the matrix model. We have also compared the results with the Polyakov loop parameter estimated from lattice QCD simulations. We have studied the temperature and momentum dependence of heavy quark drag and diffusion coefficients. It is observed that the temperature dependence of the drag coefficient is quite weak which may play a key role to understand heavy quark observables at RHIC and LHC energies. 
\end{abstract}
\maketitle
\section{Introduction}
Experimental heavy-ion collision (HIC) programs at Relativistic Heavy Ion Collider (RHIC) and  at the Large Hadron Collider (LHC) indicate the production of a liquid-like phase of the matter, having a remarkably small value of shear viscosity to entropy density ratio, $\eta/s \approx 0.1$, where the properties of the system are governed by quarks and gluons. Such a state of matter is known as quark gluon plasma (QGP)~\cite{Shuryak:2004cy, Science_Muller}. To characterize the properties of QGP, penetrating and well calibrated probes are essential. In this context, the heavy quarks (HQs)~\cite{Prino:2016cni, Andronic:2015wma, Greco:2017rro,Aarts:2016hap, Rapp:2018qla, Cao:2018ews, Dong:2019unq}, mainly charm and bottom, play a crucial role since they do not constitute the bulk part of the matter owing to their larger mass compared to the temperature created in heavy-ion collisions. Also, thermal production of heavy quarks is negligible, due to their large masses, in the QGP within the range of temperatures that can be achieved in RHIC and LHC colliding energies. 

Heavy quarks are exclusively created in hard processes which can be handled by perturbative QCD calculations~\cite{initial}, and therefore, their initial distribution is theoretically known and can be verified by experiment. They interact with the plasma constituents, the light quarks, and the gluons, but their initial spectrum is too hard to come to equilibrium with the medium. Therefore, the  high momentum heavy quarks spectrum  carry the information of their interaction with the plasma particles during the expansion of the hot and dense fireball and on the plasma properties. Since the light quark, anti-quark and gluons are thermalized, the heavy quark interaction with the light constituents leads to a Brownian motion which can be treated with the framework of a Fokker Plank equation. Thus the interaction of the heavy quark in QGP is contained in the drag and diffusion coefficients of the heavy quark. The resulting momentum distribution of the heavy mesons which depend upon the drag and diffusion coefficients get reflected in the nuclear modification factor ($R_{AA}$) which is measured experimentally. 

Initially, pQCD predicted a small nuclear suppression factor~\cite{Djordjevic:2005db,Armesto:2005mz}, $R_{AA}$, in nucleus-nucleus collisions in comparison with the proton-proton collisions. The first experiment data~\cite{stare, phenixelat, phenixelat1} on heavy quarks suggest a strong nuclear suppression factor which can not be explained within the pQCD framework.   
Several attempts~\cite{Moore:2004tg, vanHees:2005wb, vanHees:2007me,
 He:2011qa, Das:2015ana, Scardina:2017ipo, Das:2012ck, Berrehrah:2013mua, Song:2015sfa, Gossiaux:2008jv, Alberico:2011zy, 
Lang:2012cx, Xu:2017obm, Cao:2016gvr, Prado:2016szr, Nahrgang:2014vza, Plumari:2017ntm, Das:2015aga} have been made by different 
groups to study the heavy quarks interaction in QGP going beyond  pQCD to include the nonperturbative effects. Quasi-particle models enjoy considerable success in describing heavy quark dynamics in QGP~\cite{Das:2015ana, Song:2015sfa}. 

In the present study we are making a first attempt to study heavy quark transport coefficient in QGP including
 the non-perturbative effects through a background gauge field (the Polyakov loop background) and chiral condensate. 
The Polyakov loop manifests itself in the transport coefficient in two ways. Firstly, through the Debye mass that
 enter in calculating the scatterings of the heavy quark off of light thermal partons. It also enters non-trivially
 on the statistical distribution of the light partons in a non-perturbative medium. Indeed, both the effects arising
 from Polyakov loop and quark condensate are important near the transition temperature. The value of the
 normalized Polyakov loop is about half its asymptotic value at the critical temperature in different low energy
 effective models like Polyakov Nambu Jona Lasinio (PNJL) models~\cite{fukushimapnjl, rattipnjl,pedrocostapnjl}, or
 Polyakov quark meson(PQM)~\cite{bjschaefer, guptatiwari,bielich, ranjita,buballa, Abhishek:2017pkp} models. Similarly, the 
chiral condensate remains  significantly finite at temperatures around the critical temperature.  Effects of Polyakov loop
 has been studied in various contexts such as dilepton and photon production~\cite{Hidaka:2015ima}, heavy quark energy 
loss~\cite{Lin:2013efa}. Significant effects have been found by including these non-perturbative features. To estimate 
the quark masses and the Debye mass we therefore need the value of the Polyakov loop as a function of temperature. 
We do so in two different approaches. One is phenomenological in the sense that we take Polyakov loop value 
as a function of temperature from PQM model. The other approach is to take the same from lattice QCD simulations.

This paper is organized as follows, in section~\ref{formalism} we give the formalism for calculating drag and diffusion of heavy quarks by employing Boltzmann equation in soft momentum exchange between heavy quark and bulk medium \cite{Svetitsky:1987gq}. In section~\ref{mass} we recapitulate and  summarize the calculation of the Debye mass and the quark thermal mass in a Polyakov loop background as has been outlined in Refs.\cite{Hidaka:2015ima,Hidaka:2009hs}. In these calculations, we have also kept the effects of a possible finite quark mass. Such an effect can be important near the transition temperature where the light quark condensates could still be relevant. The drag and the diffusion  coefficients are evaluated in section~(\ref{results}) where we discuss their behavior as a function of temperature as well as  momentum. Finally, in section~(\ref{summary}) we summarise the results and present a possible outlook. We summarise the salient features of PQM model in Appendix(\ref{PQM}). Further, in Appendix(\ref{scatterings}), we give some details of the calculation for the square of matrix elements for the relavant $2\rightarrow 2$ processes.

\section{Formalism}
\label{formalism}
In the QGP phase, the Boltzmann equation for charm quark distribution function, neglecting any mean-field 
term, can be written as~\cite{Svetitsky:1987gq,Rapp:2009my}: 
\begin{equation}
\frac{\partial f_{HQ}}{\partial t}=
\left[\frac{\partial f_{HQ}}{\partial t}\right]_{\mathrm col},
\label{beq}
\end{equation}
where $f_{HQ}$ represents the spatially integrated non-equilibrium distribution function for heavy quark. 
The right hand side of Eq.(\ref{beq}) is the collision integral where the phase-space distribution function of the
 bulk medium appears as an integrated quantity. If we define $\omega(\zbf p,\zbf k)$ as the transition rate of collisions 
of the heavy quark with the heat bath particles (light quarks/antiquarks and
gluons)  that change the heavy quark momentum from $\zbf p$ to $\zbf p-\zbf k$, then we can write~\cite{Svetitsky:1987gq}
\begin{equation}
\left[\frac{\partial f_{HQ}}{\partial t}\right]_{col} = \int d^3k \left[ \omega(\zbf p+\zbf k,\zbf k)f_{HQ}(\zbf p+\zbf k)-
\omega(\zbf p,\zbf k)f_{HQ}(\zbf p) \right].
\label{expeq_00}
\end{equation}
The first term in the integrand represents a gain of probability
through collisions which knock the charm quark
into the volume element of momentum space at $\zbf p$ and the second
term represents the loss out of that volume element. $\omega(\zbf p,\zbf k)$ is the total
contributions coming from heavy quark scattering from gluon and light quark/anti-quark.
Furthermore, assuming the scattering processes to be dominated by small momentum transfer, we 
can expand $\omega(\zbf p+\zbf k,\zbf k)f_{HQ}(\zbf p+\zbf k)$ around $\zbf k$,
\begin{equation}
\omega(\zbf p+\zbf k,\zbf k)f_{HQ}(\zbf p+\zbf k) \approx \omega(\zbf p,\zbf k)f_{HQ}(\zbf p) +\zbf k \cdot 
\frac{\partial}{\partial \zbf p} (\omega f_{HQ}(\zbf p)) 
+\frac{1}{2}k_ik_j \frac{\partial^2}{\partial p_i \partial p_j} (\omega f_{HQ}(\zbf p)).
\label{expeq_000}
\end{equation}
The higher power of the momentum transfer, $k_i$'s, are assumed to be small~\cite{landau}.
Keeping up to the second term and substituting in Eq.(\ref{expeq_00}), we get: 
\begin{equation}
\left[\frac{\partial f_{HQ}}{\partial t}\right]_{col} = 
\frac{\partial}{\partial p_i} \left[ A_i(\zbf p)f_{HQ} + 
\frac{\partial}{\partial p_j} \lbrack B_{ij}(p) f_{HQ} \rbrack\right].
\label{expeq}
\end{equation}
Now Eq.(\ref{beq}) is reduced to Fokker-Planck equation, where the kernels 
\begin{eqnarray}
&& A_i = \int d \zbf k \omega (\zbf p,\zbf k) k_i, \nonumber\\
&&B_{ij} = \int d \zbf k \omega (\zbf p,\zbf k) k_ik_j,
\end{eqnarray}
stand for the drag and the diffusion coefficients respectively. The function $\omega(\zbf p,\zbf k)$ is given by
\begin{equation}
\omega(\zbf p,\zbf k)=g_{q,g}\int\frac{d\zbf q}{(2\pi)^3}f_{l}(\zbf q)v\sigma_{\zbf p,\zbf q\rightarrow \zbf p-\zbf k,\zbf q+\zbf k},
\end{equation}
where $f_{l}(\zbf q)$ is the thermal phase space distribution of the particles which constitute the heat bath 
which in the present case stands for  light quarks/anti-quarks and gluons,
$v=| v_p - v_q |$ is the relative velocity between the two collision partners, $\sigma$ denotes the interaction cross section  and $g_{q/g}$ is the statistical degeneracy factor for light quarks/anti-quarks and gluons. 

In particular $A_i$ and $B_{ij}$, for the (generic) process, $HQ( p) + l( q) \rightarrow HQ( p^\prime) + l( q^\prime)$ 
($l$ stands for light quarks and gluon), are given by~\cite{Svetitsky:1987gq, GolamMustafa:1997id, vanHees:2004gq, Das:2009vy}:
\begin{eqnarray}
A_i&=&\frac{1}{2E_p}\int\frac{d\zbf q}{(2\pi)^3E_q×}\int\frac{d \zbf p^\prime}{(2\pi)^3E_p^\prime×}
\int \frac{d\zbf q^\prime}{(2\pi)^3E_q^\prime×}\nonumber\\ 
&&\times\frac{1}{g_{HQ}}
\sum {|M|^2} (2\pi)^4 \delta^4(p+q-p^\prime-q^\prime)f_l(q)\nonumber\\
&&(1\pm f_l (q^\prime))[(p-p^\prime)_i] \equiv \langle \langle
(p-p^\prime)\rangle \rangle,
\label{equ1}
\end{eqnarray}
$g_{HQ}$ is the statistical degeneracy of the charm quark.
The factor $f_l(q)$ denotes the thermal phase space factor 
for the gluons and light quarks/anti-quarks in the incident channel and $1\pm f_l(q^\prime)$ is the final state Bose/Fermi enhanced/suppression phase space factor. The above expression
indicates that the drag coefficient is a measure of the thermal average of the momentum transfer, $p-p^\prime$, weighted by the elastic heavy quark-bulk interaction through the square of the invariant amplitude, ${\mid M\mid^2}$. 

Similar, heavy quark diffusion coefficients can be defined as:
\begin{eqnarray}
B_{ij}&=&\frac{1}{2E_p}\int\frac{d \zbf q}{(2\pi)^3E_q×}\int\frac{d\zbf p^\prime}{(2\pi)^3E_p^\prime×}
\int \frac{d\zbf q^\prime}{(2\pi)^3E_q^\prime×}\nonumber\\ 
&&\times\frac{1}{g_{HQ}}
\sum  {|M|^2} (2\pi)^4 \delta^4(p+q-p^\prime-q^\prime)f_l(q)\nonumber\\
&&(1\pm f_l(q^\prime))\bigg[\frac{1}{2×}(p-p^\prime)_i(p-p^\prime)_j\bigg] \equiv \langle \langle
(p-p^\prime)_i(p-p^\prime)_j\rangle \rangle.
\label{equ11}
\end{eqnarray}
From the above expression it is clear that the diffusion coefficient is a measure of the thermal average 
of the square of momentum transfer weighted by the elastic heavy quark-bulk interaction through the square of the invariant amplitude,
${|M|^2}$. Since $A_i$ and $B_{ij}$ depend only on the vector $\zbf p$, we may write~\cite{Svetitsky:1987gq}:
\begin{eqnarray}
A_i=p_iA,
\end{eqnarray}
\begin{eqnarray}
B_{ij}=\bigg(\delta_{ij}-\frac{p_ip_j}{\zbf p^2×}\bigg)B_0+\frac{p_ip_j}{\zbf p^2×}B_1,
\end{eqnarray}
where,
\begin{eqnarray}
A=p_iA_i/\zbf p^2=\langle \langle 1\rangle \rangle-
\frac{\langle \langle \zbf p.\zbf p\prime \rangle \rangle}{\zbf p^2},
\label{drag_cof}
\end{eqnarray}
\begin{eqnarray}
B_0=\frac{1}{2×}\bigg(\delta_{ij}-\frac{p_ip_j}{\zbf 2p^2×}\bigg)B_{ij}=\frac{1}{4×}
\bigg[\langle \langle \zbf p\prime^2\rangle \rangle-\frac{\langle \langle (\zbf p.\zbf p\prime)^2 \rangle \rangle}{\zbf p^2}\bigg],
\label{diff_cof}
\end{eqnarray}
\begin{eqnarray}
B_1=\frac{p_ip_j}{\zbf p^2×}B_{ij}=\frac{1}{2×}\bigg[\frac{\langle \langle (\zbf p.\zbf p\prime)^2 \rangle \rangle}{\zbf p^2}-2\langle 
\langle \zbf p.\zbf p\prime \rangle \rangle+\zbf p^2\langle \langle 1\rangle \rangle\bigg].
\end{eqnarray}
The integrals appearing in the above  equations can be further simplified by solving the kinematics in the center of mass frame of the colliding particles and both the drag and diffusion coefficients can be defined from a single expression:
\begin{eqnarray}
\langle \langle \Gamma(p^\prime)\rangle \rangle &=&\frac{1}{512\pi^4×}\frac{1}{E_p} 
\int_{0}^{\infty} \int_{-1}^{1}d(cos\theta_{cm})
\int_0^{2\pi}d\phi_{cm}
\frac{q^2 dq d(cos\chi)}{E_q}
{f}(q)(1\pm f(q^\prime))\nonumber\\
&\times&\frac{\lambda^{\frac{1}{2}}(s,m_C^2,m_q^2)}{\sqrt{s}} 
\frac{1}{g_{HQ}} \sum  {|M|^2}{{\Gamma}}(p^\prime),
\label{transport11}
\end{eqnarray}
with an appropriate choice of ${{\Gamma}}(p^\prime)$. As in Ref.\cite{Svetitsky:1987gq}, we shall consider $2\rightarrow 2$ processes which involve Coulomb scattering i.e., $q Q\rightarrow qQ$ through gluon exchange and Compton scattering of gluon and heavy quark i.e., $gQ \rightarrow gQ$. In the  present work, we shall estimate the scattering amplitudes in the background of Polyakov loop. This makes the square of the corresponding matrix element as well as distribution function dependent on the color indices (see e.g., Eqs.(\ref{quarkdist}) and (\ref{gluondist})). Therefore, the expression for $\langle \langle \Gamma(p^\prime)\rangle \rangle$ becomes 
\begin{eqnarray}
\langle \langle \Gamma(p^\prime)\rangle \rangle &=&\frac{1}{512\pi^4×}\frac{1}{E_p} 
\int_{0}^{\infty} \int_{-1}^{1}d(cos\theta_{cm})
\int_0^{2\pi}d\phi_{cm}
\frac{q^2 dq d(cos\chi)}{E_q}\frac{\lambda^{\frac{1}{2}}(s,m_C^2,m_q^2)}{\sqrt{s}} 
\frac{1}{g_{HQ}}\nonumber\\
&\times& \bigg(\sum _{abef}f(q)_{e}(1- f(q^\prime)_{f}){|M_{C}|_{abef}^2}+\sum _{abefgh}f(q)_{ef}(1+ f(q^\prime)_{gh}){|M_{Cm}|_{abefgh}^2}\bigg){{\Gamma}}(p^\prime),
\label{transport111}
\end{eqnarray}
where ${|M_{C}|_{abef}^2}$ is matrix element squared for $q^{a}Q^{b}\rightarrow q^{e}Q^{f}$ with $ab(ef)$ as initial(final) quark color indices  and ${|M_{Cm}|^2}_{abefgh}$ is matrix element squared for $g^{ef}Q^{b} \rightarrow g^{gh}Q^{a}$ scatterings with $ef,a(gh,b)$ as initial(final) gluon and quark color indices. Here the color indices $a,b,e,f,g,h=1,2,3$ are in fundamental representation. Furthermore, in Eq.(\ref{transport111}), $\lambda(x,y,z)=x^2+y^2+z^2 -2xy-2yz-2zx$ is the triangular function. $E_p$ and $m_C$ are the heavy quark energy and mass, respectively. $E_q$ is the energy of the light quark/gluon. $s$ is the Mandelstam variable. To compute the heavy quark transport coefficient, one needs, therefore, the heavy quark-light quark/gluon scattering matrix along with the thermal distribution functions, the mass of light quarks and gluons and Debye screening mass. The divergence in t-channel diagram here is regulated by a Debye mass ~\cite{Svetitsky:1987gq}.

In literature several attempts have been made, over the years, to compute the heavy quark drag and diffusion 
coefficients in QGP within different models. A recent study indicates that non-perturbative contributions are essential
for the simultaneous description of heavy quarks $R_{AA}$ and $v_2$~\cite{Das:2015ana}. Quasi-particles model is a way to take into account the non-perturbative effect. This can be done in a number of possible ways which differ in how the effects of QCD 
interactions are modeled. To study the  heavy quark transport properties in QGP, the quasi-particle 
approaches~\cite{Das:2015ana, Song:2015sfa} that have been recently used in literature include the interaction in 
the effective masses of the light quark and gluons. In these quasi-particle models strong coupling constant~\cite{Plumari:2011mk}, 
$g(T)$, is the only free parameter which can be obtained by making a fit of the energy density obtained by lattice QCD calculations. The main feature of these quasi-particle approach is that the resulting coupling is significantly stronger than the one obtained from pQCD running coupling particularly near the quark-hadron transition temperature ($T_c$). In this present study we adopted a different model to include the non-perturbative effects. The statistical distribution function, thermal mass and Debye mass have been obtained in presence of a non-trivial Polyakov loop background. In the following section we attempt to estimate quark thermal mass and the Debye mass in Polyakov loop background.

\section{Thermal and Debye masses in Polyakov loop background}
\label{mass}
In this section, we shall estimate the non-perturbative Debye screening mass and quark thermal mass in a nontrivial Polyakov 
loop background to be used in the estimation of the drag and diffusion coefficients using Eqs.(\ref{drag_cof}) and (\ref{diff_cof}). Such a calculation has been performed in detail in Refs.\cite{Hidaka:2015ima,Lin:2013efa, Hidaka:2009ma,Hidaka:2009hs} using a matrix model for semi-qgp and used for estimating shear viscosity to entropy ratio as well as to dilepton and photon production and energy loss of heavy quark in the medium. We recapituate the salient features of such a calculation including also the possible effects from a finite mass of the light quarks which can arise from a nonvanishing scalar quark- antiquark condensate. 

Polyakov loop is a particular case of the Wilson loop where the gluon field is time-like. The background gauge field
can be taken as  a constant diagonal matrix $A_\mu^{ab}=\delta_{\mu 0}\delta^{ab} Q^{a}/g$, where, the color 
index $a$ is not summed  and  $g$ is  the  gauge field coupling constant. The Wilson line in the temporal direction is given by 
\begin{equation}
P={\cal P} \exp\bigg(ig\int_0^{\beta} d\tau A_0(x_0,\bf x)\bigg),
\end{equation}
where, ${\cal P}$ denotes path ordering in the imaginary time, with $\tau$ being the imaginary time $\tau:0\rightarrow \beta $. 
In the mean field level, neglecting the fluctuations and with the choice of the time-independent constant background field, the path ordering becomes irrelevant and one can perform the integration over the imaginary time leading to $P=\exp(ig\beta A_0)$. The trace of the Wilson line is the Polyakov loop $\phi$ given as
\begin{equation}
\phi(Q)=\frac{1}{3}\sum_{a=1}^3\exp(i\beta Q^a).
\end{equation}
In a SU(N) gauge group the vector potential $A_0$ is traceless so the sum over all the Q's 
vanishes i.e., $\sum_{a}Q^{a}=0$, for $SU(3)$, one can parameterize $Q^a=2 \pi T(-q,0,q)$ ,
 where we have introduced a dimensionless Polyakov loop dependent parameter ``$q$"~\cite{Lin:2013efa}, so that. 
\begin{equation}
\phi=\frac{1}{3}\left(1+2 \cos 2\pi q\right).
\label{qexpr}
\end{equation}
Physically, such a nontrivial background field $A_{0}$ can be thought of as an imaginary chemical potential~\cite{Dumitru:2010mj}. 
The thermal distribution function for the quarks/anti-quarks and the gluons are given respectively by~\cite{Hidaka:2015ima}
\begin{equation}
f_a(E)=\frac{1}{e^{\beta(E-i Q^a)}+1}, \hspace{0.5cm} \tilde{f}_{a}(E)=\frac{1}{e^{\beta(E+i Q^{a})}+1},
\label{quarkdist}
\end{equation}  
\begin{equation}
f_{a b}(E)=\frac{1}{e^{\beta(E-i(Q^a-Q^b))}-1}.
\label{gluondist}
\end{equation}
Let us note  that the quark distribution function involves only one color index because these are represented in fundamental representation. For gluons, the adjoint representation leads to two fundamental indices. For three colors, the color averaged statistical distribution function of the gluons becomes
\begin{equation}
f_g(E)=\frac{1}{3^2}\sum_{a,b=1}^{3}f_{a b}(E)=\frac{1}{9}\bigg(\frac{3}{e^{\beta E}-1}+\frac{e^{\beta E}(6 \phi-2)-4}{1+e^{2\beta E}+e^{\beta E}(1-3\phi)}+\frac{e^{\beta E}(9\phi^2-6\phi-1)-2}{1+e^{2\beta E}+e^{\beta E}(1+6\phi-9\phi^2)}\bigg).
\end{equation}
Similarly, for three colors, the color averaged distribution functions of the quark/anti-quark is
\begin{equation}
f_{q/\bar{q}}(E)=\frac{1}{3}\sum_{a=1}^{3} f_a(E)=\frac{1}{3}\sum_{a=1}^{3} \tilde{f}_a(E)=\frac{\phi e^{-\beta E}+2 \phi e^{-2 \beta E}+ e^{-3 \beta E}}{1+3 \phi e^{-\beta E}+3 \phi e^{-2 \beta E}+e^{-3 \beta E}}.
\label{avgfunq}
\end{equation}
It may be noted that for pure gluon case, $\phi=1$ in the confined phase and $\phi=0$ in the deconfined phase. This leads to the gluon distribution function 
\begin{equation}
f_g(E)=\frac{1}{e^{3 \beta E}-1},
\end{equation}
in the confined phase and
\begin{equation}
f_g(E)=\frac{1}{e^{ \beta E}-1},
\end{equation} 
in the deconfined phase. In the presence of quarks, one does not have a rigorous order parameter 
for deconfinement, however in $\phi=0$ case the color averaged quark/anti-quark distribution reduces to 
\begin{equation}
f_{q/\bar{q}}(E)=\frac{1}{e^{3 \beta E}+1}
\end{equation}
so that quark are suppressed statistically. In the perturbative limit i.e., $\phi=1$ it becomes
\begin{equation}
f_{q/\bar{q}}(E)=\frac{1}{e^{\beta E}+1}.
\end{equation}
The color averaged distribution function of quark/anti-quark as given in Eq.(\ref{avgfunq}) is exactly the same as that 
in PQM model within mean field approximation~\cite{Abhishek:2017pkp}. For the computation of Debye and thermal mass, we use
 double line notation~\cite{tHooft:1973alw,Cvitanovic:1976am} which is convenient for large $N_c$ calculations. 
 For $SU(N)$ gauge group, the generators $\lambda^A$ satisfy the following relation~\cite{Hidaka:2009hs}
\begin{equation}
Tr(\lambda^A \lambda^B)=\frac{1}{2} \delta^{A B},
\end{equation} 
where $A$ and $B$ are adjoint indices and takes the values $A,B=1,2,3,..,N^2-1$. Each adjoint indices can be denoted by a pair of fundamental indices. For double line notation, the quantity that we need here is the projection operator, with adjoint indices it is written as
\begin{equation}
\mathcal{P}^{k l}_{m n}=\delta^{k}_{m}\delta^{l}_{n}-\frac{1}{N} \delta^{k l}\delta_{m n}.
\end{equation}
In the calculation of quark and gluon self energies, one needs the vertices for quark-antiquark-gluon($q \bar{q}g$) interaction, which is proportional to the generators. In the double line notation, the generators in the fundamental representation are written as
\begin{equation}
t^{a b}_{c d}=\frac{1}{\sqrt{2}}\mathcal{P}^{a b}_{c d}.
\end{equation}
Here upper pair $ab$ denotes the adjoint index while the lower pair $c d$ denotes the components of this matrix in the fundamental representation. Similarly, the triple-gluon vertex is proportional to structure constants which in the double line notation can be written as
\begin{equation}
f^{(k l, m n, a b)}=\frac{i}{\sqrt{2}}(\delta^{k n}\delta^{m b}\delta^{a l}-\delta^{k b}\delta^{m l}\delta^{a n}).
\label{structure}
\end{equation}
\subsection{Quark loop contribution to Debye mass}
Generally, Debye mass ($m_D$) is defined through the pole of effective propagator in the static limit 
i.e., $\omega=0,\vec{p}\rightarrow 0 $ and is related to the time-like component of gluon self-energy 
$\Pi_{44}(\omega=0,\vec{p}\rightarrow 0 )$~\cite{Bellac:2011kqa}. It turns out that, in the presence of 
a static background field, apart from the usual $T^2$ dependent term similar to as in perturbative HTL calculations, 
there is an additional $T^3$ dependent contribution to the gluon self energy. The later component arises because 
the background field induces a color current which couples to the gluon. While the $T^2$ dependent 
term in $\Pi_{\mu \nu}$ is transverse (i.e., $P^{\mu}\Pi^{\mu \nu}(P)=0$), the $T^3$ dependent term is not and spoils the transversality relation which
 is required for the gauge invariance. Therefore, one needs an additional contribution which may 
be of non-perturbative origin to the gluon self energy to cancel such a term. Similar to Ref.\cite{Hidaka:2009ma},
 we assume that such a term exists and cancels this undesirable $T^3$ term. Under these assumptions, 
the Polyakov loop dependent resummed propagator can be written as~ \cite{Hidaka:2009ma}   
\begin{equation}
D{\mu \nu ; a b c d}=P^{L}_{\mu \nu} \frac{k^2}{K^2} D^{L}_{a b c d}(K)+P^{T}_{\mu \nu}D^{T}_{a b c d}(K),
\label{propef}
\end{equation}
where $P^T_{\mu \nu}=g_{\mu i}\bigg(-g^{i j}-\frac{k^{i} k^{j}}{K^2}\bigg)g_{j\nu}$ and $P^L_{\mu \nu}=-g_{\mu \nu}+\frac{k_{\mu}k_{\nu}}{K^2}-P^{T}_{\mu \nu}$ respectively are  the longitudinal and the transverse projection operators and are defined as
\begin{equation}
D^{L}_{\mu \nu ; a b c d}(K)=\bigg(\frac{i}{K^2-F}\bigg)_{a b c d},
\end{equation}
\begin{equation}
D^{T}_{\mu \nu ; a b c d}(K)=\bigg(\frac{i}{K^2-G}\bigg)_{a b c d},
\end{equation}
where 
\begin{equation}
F=-2 m^2\bigg(1-\frac{x}{2}\ln\bigg(\frac{x+1}{x-1}\bigg)\bigg),
\end{equation}
\begin{equation}
G= m^2\bigg(x^2+\frac{x(1-x^2)}{2}\ln\bigg(\frac{x+1}{x-1}\bigg)\bigg),
\end{equation}
with $x=\frac{k_{0}}{k}$ and $m^2=(m^2)_{a b c d}$ is the thermal mass of the gluon. Under the assumptions taken here, it is clear that the pole (F) of the longitudinal propagator can be related to $\Pi_{44}$ component of gluon self energy. Furthermore, in the static limit, this term can be defined as Debye mass~\cite{Hidaka:2009hs}. 

\begin{figure}[tbh]
\includegraphics[width=12cm]{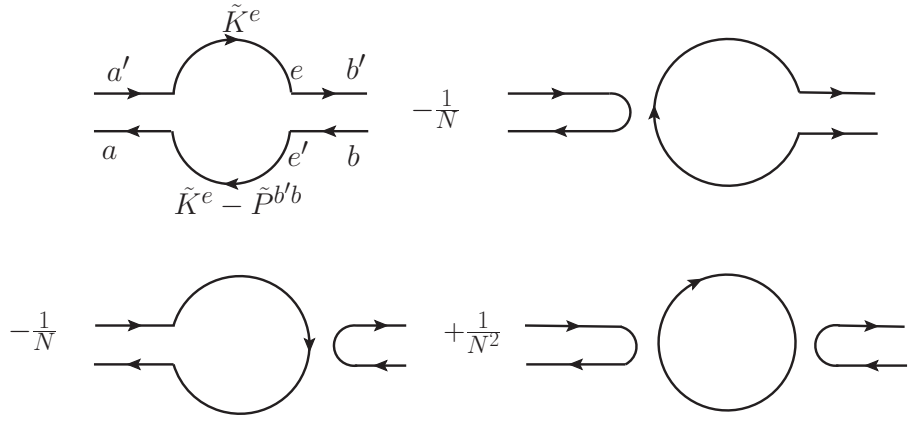}
\caption{Quark loop of gluon self energy in double line notation.}
\label{quarkloop}
\end{figure}

In this work, we shall focus only on the time like component of the gluon self energy with the assumption that $T^3$ dependent term
is cancelled.  For massless quarks, Debye mass has already been computed in Ref.\cite{Hidaka:2009hs}. We include here
the effect of finite constituent quark mass in the quark loop contribution to the Debye mass. We work in the imaginary 
time formalism of thermal field theory for evaluating the corresponding diagrams. In this formalism, 
because of the boundary conditions of imaginary time, the energy of a fermion $p_4$ is an odd multiple of $\pi T$ while that for a boson is an even multiple of $\pi T$. For calculating the Debye mass, we first evaluate the quark loop in the gluon self-energy for which the corresponding diagram is shown in Fig.(\ref{quarkloop}), where the loop momentum 
four vector is written as $\tilde{K}^e_{\mu}= (K+\tilde{Q}_e)_{\mu}=(\omega_n+\tilde{Q}^e, \zbf{k})$ with $\tilde{Q}_e=Q_e+\pi T$.
In t'hooft double line notation, the polarization tensor can be written as
\begin{equation}
\Pi_{\mu \nu;b' b a a'}^q(P,Q,m)=g^2 N_f t^{a a'}_{e e'}t^{b b'}_{e' e} \int \frac{d^4K}{(2\pi)^4}Tr_D[\gamma_{\mu}(\slashed{\tilde{K}}_e-\slashed{\tilde{P}}_{b b'})\gamma_{\nu}\slashed{\tilde{K}}_e+m^2 \gamma_{\mu}\gamma_{\nu}]{\Delta}(K){\Delta}(P-K),
\label{self1}
\end{equation}
where $aa',bb'(e,e')$ are color indices of gluons (quark/antiquark), $N_f$ is quark flavor number and ${\Delta}(K)^{-1}
={(\omega_n+\tilde{Q}_e)^2+\zbf k^2+m^2}$, ${\Delta}(P-K)^{-1}={(\omega-\omega_n+{Q}_{b b'}-\tilde{Q}_e)^2+(\zbf p-\zbf k)^2+m^2}$ 
with $Q_{b b'}=Q_b-Q_{b'}$, $E_k=\sqrt{\zbf k^2+m^2}$, $E_q=\sqrt{(\zbf p-\zbf k)^2+m^2}$, $\omega_n=(2 n+1) \pi T$ and $P_4=\omega $. 
$Tr_{D}$ is trace in Dirac space and $Q_i$ is the diagonal matrix in color space which is given as $Q_a=(-2 \pi T q,0,2 \pi T q)$ 
and $q$ is related to the Polyakov loop expectation value as given in Eq.(\ref{qexpr}). Here, we take hard thermal loop (HTL) 
approximation and also assume that $m \ll T$. Thus, taking HTL limit and the trace over Dirac space, Eq.(\ref{self1}) reduces to
\begin{equation} 
\Pi_{\mu \nu;b' b a a'}^q(P,Q,m)=g^2 N_f t^{a a'}_{e e'}t^{b b'}_{e' e} \int \frac{d^4K}{(2\pi)^4}[8 (K+\tilde{Q}_e)_{\mu}(K+\tilde{Q}_e)_{\nu}-4(K+\tilde{Q}_{e})^2 \delta_{\mu \nu}-4 m^2 \delta_{\mu \nu}]{\Delta}(K){\Delta}(P-K).
\label{self}
\end{equation}
As we are interested in calculating Debye mass for which we need  time-like component ($\Pi_{44}$) of the gluon self-energy. 
So from here onwards, we shall proceed with this term. For this purpose, we write the integration in Eq.(\ref{self}) as
$\int \frac{d^4K}{(2 \pi)^4}=T \sum\limits_{n=-\infty}^{\infty} \int \frac{d{\zbf{k}}}{(2\pi)^3}; k_4\equiv\omega_n=2n\pi T$. Simplifying Eq.(\ref{self}), we have
\begin{equation} 
\Pi_{4 4;b' b a a'}^q(P,Q,m)=4g^2 N_f t^{a a'}_{e e'}t^{b b'}_{e' e} \int \frac{d\zbf k}{(2\pi)^3}T\sum_{n}[(-2 k^2- m^2){\Delta}(K){\Delta}(P-K)+{\Delta}(P-K)].
\label{fself}
\end{equation}
The frequency sums in Eq.(\ref{fself}) over discrete Matsubara frequencies are somewhat involved but can be performed routinely leading to 
\begin{eqnarray}\nonumber 
T\sum_{n=-\infty}^{\infty}\Delta(K)\Delta(P-K)&=&
\frac{1}{4 E_k E_q}\bigg(\frac{f(E_q+i Q2+i \omega)-f(E_k-i Q1)}{E_k-E_q+i(Q1+Q2+\omega)}
+\frac{1+f(E_k-i Q1)-f(E_q-i Q2-i\omega)}{E_k+E_q-i (Q1+Q2+\omega)}\nonumber\\
&+&\frac{f(E_k+i Q1)-f(E_q-i Q2-i\omega)}{E_q-E_k+i(Q1+Q2+\omega)}
+\frac{1+f(E_k+i Q1)-f(E_q+i Q2+i\omega)}{E_k+E_q+i(Q1+Q2+\omega)}\bigg),
\label{frq1} 
\end{eqnarray}
\begin{equation}
T\sum_{n=-\infty}^{\infty}\Delta(P-K)=-\frac{1+f(E_q+i Q2+i\omega)+f(E_q-i Q2-i\omega)}{2 E_q},
\label{frq2}
\end{equation}
where $Q2=Q_{b b'}-\tilde{Q}_e$, $Q1=\tilde{Q}_e$ and $f(E \pm iQ)$ is Bose-Einstein distribution function.
 In Eqs.(\ref{frq1}) and (\ref{frq2}), the term which is independent of distribution function is the vacuum contribution
 which can be dropped when one considers the medium dependent terms only.  First and third term in Eq.(\ref{frq1}) contribute 
to the $T^3$ dependent term. Such a term exists only in the presence of a background gauge field in the HTL 
approximations~\cite{Hidaka:2009hs}. As mentioned earlier, this term spoils the transversality condition and we shall not consider 
this undesirable contribution. Furthermore, the $T^2$ dependent contributions are given by second and fourth term 
of Eq.(\ref{frq1}) as well as by the medium dependent term in Eq.(\ref{frq2}). In the static limit, the time 
like component of the gluon self-energy can be written as
\begin{equation}
\Pi_{4 4;b' b a a'}^q(Q,m)|_{(\omega=0,\vec{p}\rightarrow 0 )}=-4g^2 N_f t^{a a'}_{e e'}t^{b b'}_{e' e} [2 I_{1}(m,\tilde{Q}_e,Q_{b b'}-\tilde{Q}_e)+I_{2}(m,\tilde{Q}_e,Q_{b b'}-\tilde{Q}_e)+I_{3}(m,Q_{b b'}-\tilde{Q}_e)],
\end{equation}
where
\begin{equation}
I_{1}(m,\tilde{Q}_e,Q_{b b'}-\tilde{Q}_e)=\frac{T^2}{16 \pi^2}\int{\frac{x^4 dx}{{(x^2+y^2)^{\frac{3}{2}}}} }\bigg(f(x,y,iq1)+f(x,y,-iq1)-f(x,y,iq2)-f(x,y,-iq2)\bigg),
\label{eqn9}
\end{equation}
\begin{equation}
I_{2}(m,\tilde{Q}_e,Q_{b b'}-\tilde{Q}_e)=\frac{m^2}{16 \pi^2}\int{\frac{x^2 dx}{{(x^2+y^2)^{\frac{3}{2}}}} }\bigg(f(x,y,iq1)+f(x,y,-iq1)-f(x,y,iq2)-f(x,y,-iq2)\bigg),
\label{I2}
\end{equation}
\begin{equation}
I_{3}(m,Q_{b b'}-\tilde{Q}_e)=\frac{T^2}{4 \pi^2}\int{\frac{x^2 dx}{\sqrt{x^2+y^2}} }\bigg(f(x,y,iq2)+f(x,y,-iq2)\bigg),
\label{eqn10}
\end{equation} 
where we have defined the dimensionless variables $x=\beta k$, $y=\beta m$ and $q1=\beta Q1$. Further, 
$f(x,y,iq)$'s are the Bose distribution functions in terms of these dimensionless variables as e.g.,
\begin{equation}
f(x,y,iq)=\frac{1}{\exp(\sqrt{x^2+y^2}+iq)-1}.
\end{equation}
Also note that although distribution function is a complex quantity, the functions  $I_{1}(m,\tilde{Q}_e,Q_{b b'}-\tilde{Q}_e), I_{2}(m,\tilde{Q}_e,Q_{b b'}-\tilde{Q}_e)$ 
and $I_{3}(m,Q_{b b'}-\tilde{Q}_e)$ are real functions. With further simplification, $\Pi_{44}$ can be written as
\begin{equation} 
\Pi_{4 4;b' b a a'}^q(Q,m)|_{(\omega=0,\vec{p}\rightarrow 0 )}=-g^2 N_f t^{a a'}_{e' e}t^{b b'}_{e e'}
\frac{T^2}{4 \pi^2}\bigg[2(\mathfrak{D}(q1,y)-\mathfrak{D}(q2,y))+4\mathfrak{F}(q2,y)+2y^2\mathfrak{B}(Q2,y)\bigg],
\label{pi44}
\end{equation}
where the dimensionless real functions $\mathfrak{D}, \mathfrak{F}$ and $\mathfrak{B}$  are 
\begin{equation}
\mathfrak{D}(q,y)=\int{\frac{x^4 dx}{({x^2+y^2})^{\frac{3}{2}}} }\bigg(f(x,y,iq)+f(x,y,-iq)\bigg),
\label{dqy}
\end{equation}
\begin{equation}
\mathfrak{B}(q,y)=\int{\frac{x^2 dx}{({x^2+y^2})^{\frac{3}{2}}} }\bigg(f(x,y,iq)+f(x,y,-iq)\bigg),
\label{bqy}
\end{equation}
\begin{equation}
\mathfrak{F}(q,y)=\int{\frac{x^2 dx}{\sqrt{x^2+y^2}} }\bigg(f(x,y,iq)+f(x,y,-iq)\bigg).
\label{funfF}
\end{equation}
In the limiting case of vanishing quark masses i.e. $y=0$, the function $\mathfrak{B}(q,y)$ do not contribute to $\Pi^q_{44}$ as
it is multiplied by a $y^2$ term while the functions $\mathfrak{D}(q,y=0)$ and $\mathfrak{F}(q,y=0)$ become equal and can be written in terms of Polylog functions $Li_2(z)$ as
\begin{equation}
\mathfrak{F}(q,y=0)=\mathfrak{D}(q,y=0)=\int{dx x }\bigg(f(x,y=0,iq)+f(x,y=0,-iq)\bigg)\equiv Li_2(iq)+Li_2(-iq).
\label{polylog}
\end{equation}
The Polylog function $Li_2(z)$ can also be written in terms of Clausen functions $Cl_2(z)$ e.g.
\begin{equation}
Li_2(i 2\pi q )= \frac{\pi^2}{6} (1-6q+6q^2)+iCl_2(2\pi q),
\end{equation}
that has been used in Ref.\cite{Hidaka:2009hs}. In the present investigation, however, we will keep the effect of masses
in Eqs(\ref{dqy}), (\ref{bqy}), (\ref{funfF}) and integrate it numericaly to estimate the Debye mass. Generators appearing in the right side of Eq.(\ref{pi44}) can be simplified by using projection operators, so that the product of two generators becomes
\begin{eqnarray} 
t^{a a'}_{e' e}t^{b b'}_{e e'}&=&\frac{1}{2}\bigg[\delta^{b e}\delta^{b' e'}\delta^{a' e}\delta^{a e'}-\frac{1}{N}\bigg(\delta^{b b'}\delta^{e e'}\delta^{a' e}\delta^{a e'}+\delta^{b e}\delta^{b' e'}\delta^{a a'}\delta^{e' e}\bigg)+\frac{1}{N^2}\delta^{b b'}\delta^{e e'}\delta^{a a'}\delta^{e' e}\bigg].
\label{prod}
\end{eqnarray}
Note that $\Pi_{44}$ depends on the color of quark and gluon and has $a,b,a',b'$ as free color indices. So we need to sum over other repeated color indices (i.e., $e,e'$) which can be done by contracting color indices of Eq.(\ref{prod}) with that of Eq.(\ref{pi44}). Using Eq.(\ref{prod}) along with Eq.(\ref{pi44}) and summing over contracted color indices, gluon self energy can be written as
\begin{eqnarray}
\Pi_{4 4;b' b a a'}^q(Q,m)|_{(\omega=0,\vec{p}\rightarrow 0 )}&=&-g^2 N_f\frac{T^2}{4 \pi^2}\bigg[\delta_{ab}\delta_{a'b'}\bigg(\mathfrak{D}(\tilde{Q}_b,y)-\mathfrak{D}(\tilde{Q}_{b'},y)+2\mathfrak{F}(\tilde{Q}_{b'},y)+y^2\mathfrak{B}(\tilde{Q}_{b'},y)\bigg)\nonumber \\
&-&\frac{1}{N}\bigg( \mathfrak{D}(\tilde{Q}_{b'},y)+\mathfrak{D}(\tilde{Q}_{a'},y)+ \mathfrak{F}(\tilde{Q}_{b'},y)+\mathfrak{F}(\tilde{Q}_{a'},y)+2 y^2\mathfrak{B}(\tilde{Q}_{a'},y)\nonumber\\
&+& 2 y^2\mathfrak{B}(\tilde{Q}_{b'},y)\bigg)\delta_{aa'}\delta_{bb'}+\frac{1}{N^2}\sum_{e}\bigg( \mathfrak{D}(\tilde{Q}_{e},y)+ \mathfrak{F}(\tilde{Q}_{e},y)\nonumber \\
&+&2 y^2 \mathfrak{B}(\tilde{Q}_{e},y)\bigg)\delta_{aa'}\delta_{bb'}\bigg].
\label{selffer1}
\end{eqnarray}	

\subsection{Gluon contribution to Debye mass}
\begin{figure}[tbh]
\includegraphics[width=12cm]{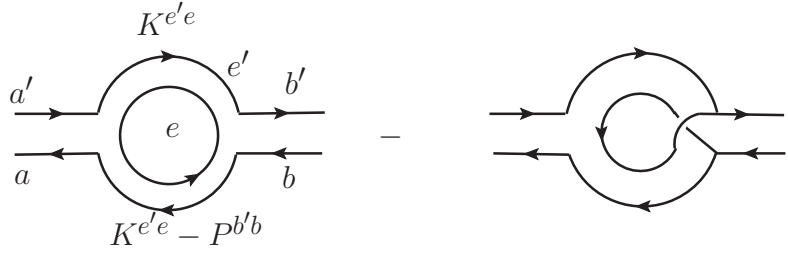}
\caption{Gluon loop in gluon self energy in double line notation}
\label{gluonloop}
\end{figure}

Gluon loop contribution to the gluon self energy  has already been evaluated in Ref.\cite{Hidaka:2009hs}. For the sake of completeness, we recapitulate the results here. Gluon loop diagram with tri-gluon vertex is  shown in Fig.(\ref{gluonloop}). 
In the HTL approximation, the sum of gluon loop, four gluon vertex and ghost loop contribution to the gluon self energy can be written as
\begin{equation} 
\Pi_{\mu \nu;b' b a a'}^{gl}(P,Q)=g^2 f^{(b' b,e e', g h)}f^{(a a',e' e, h g)} \int \frac{d^4 K}{(2 \pi)^4} [4 K_{\mu e' e}K_{\nu e' e}-2 K_{e' e}^2 \delta_{\mu \nu}]\Delta(K)\Delta(P-K).
\end{equation}
As explained earlier, the time like component of the self energy is needed for the Debye mass which can be written as
\begin{equation} 
\Pi_{44;b' b a a'}^{gl}(P,Q)= g^2 f^{(b' b,e e', g h)}f^{(a a',e' e, h g)} \int \frac{d \zbf{k}}{(2 \pi)^3}\sum_{n} T[2\Delta(P-K)-4 k^2 \Delta(K)\Delta(P-K)],
\label{gluself}
\end{equation}
where $\Delta(K)^{-1}={(\omega_n+Q_{e' e})^2 }$ and $\Delta(P-K)^{-1}={(\omega-\omega_n+Q_{b' b}-Q_{e' e})^2+E_q^2}$. 
Here $Q1=Q_{e' e}$ and $Q2=Q_{b' b}-Q_{e' e}$.  Similar to quark loop, we shall not consider $T^3$ dependent term here 
and the summation over discrete Matsubara frequencies are same as in Eqs.(\ref{frq1}) and (\ref{frq2}). Using these summations and taking static limit, the $T^2$ dependent contribution to gluon self energy can be written as
\begin{equation}
\Pi_{44;b' b a a'}^{gl}(Q)|_{(\omega=0,\vec{p}\rightarrow 0 )}=- \frac{g^2 T^2}{4 \pi^2} f^{(b' b,e e', g h)}f^{(a a',e' e, h g)}[3\mathfrak{H}(Q_{b' b}-q_{e' e})+\mathfrak{H}(Q_{e' e})],
\label{pi441}
\end{equation}
where 
\begin{equation}
\mathfrak{H}(Q)=\int x dx (f(x,iq)+f(x,-iq))\equiv Li_2(iq)+Li_2(-iq).
\end{equation}
Same as in the case of quark loop, for gluon loops, gluon self energy depends on the color of the gluon, and these color indices are free. Other repeated color indices can be summed by using Eq.(\ref{structure}) for structure constant. Thus Eq.(\ref{pi441}) becomes
\begin{equation}
\Pi_{44;b' b a a'}^{gl}(Q)|_{(\omega=0,\vec{p}\rightarrow 0 )}= \frac{g^2 T^2}{8 \pi^2} [4(\mathfrak{H}(Q_{b a})+\mathfrak{H}(Q_{a b}))\delta^{b' b}\delta^{a' a}-2(3\mathfrak{H}(Q_{b e})+\mathfrak{H}(Q_{b' e}))\delta^{a' b'}\delta^{a b}].
\label{selfglu}
\end{equation}
To get the total Debye mass we need to add both the contribution which are given in Eqs.(\ref{selffer1}) and (\ref{selfglu}). Taking both the contributions into account, Debye mass can be given as
\begin{equation}
(m_D^2)_{b'baa'}=-\Pi_{44;b'baa'}^{q}(m)|_{(\omega=0,\vec{p}\rightarrow 0 )}-\Pi_{44;b'baa'}^{gl}(Q)|_{(\omega=0,\vec{p}\rightarrow 0 )},
\end{equation}
leading to
\begin{eqnarray}
(m_D^2)_{b'b a a'}&=&\frac{g^2T^2}{4 \pi^2}\bigg[N_f\bigg(\delta_{ab}\delta_{a'b'}\bigg(\mathfrak{D}(\tilde{Q}_b,y)-\mathfrak{D}(\tilde{Q}_{b'},y)+2\mathfrak{F}(\tilde{Q}_{b'},y)+y^2\mathfrak{B}(\tilde{Q}_{b'},y)\bigg)\nonumber \\
&-&\frac{1}{N}\bigg(\mathfrak{D}(\tilde{Q}_{b'},y)+\mathfrak{D}(\tilde{Q}_{a'},y)+ \mathfrak{F}(\tilde{Q}_{b'},y)
+\mathfrak{F}(\tilde{Q}_{a'},y)+2 y^2\mathfrak{B}(\tilde{Q}_{a'},y)+2	 y^2\mathfrak{B}(\tilde{Q}_{b'},y) \bigg)\delta_{aa'}\delta_{bb'}\nonumber\\
&+&\frac{1}{N^2}\sum_{e}\bigg( \mathfrak{D}(\tilde{Q}_{e},y)+ \mathfrak{F}(\tilde{Q}_{e},y)+2 y^2 \mathfrak{B}(\tilde{Q}_{e},y)\bigg)\delta_{aa'}\delta_{bb'}\bigg)+\bigg(3\mathfrak{H}(Q_{b e})+\mathfrak{H}(Q_{b' e})\bigg)\delta^{a b}\delta^{a' b'}\nonumber\\
&-&\bigg(2(\mathfrak{H}(Q_{b a})-\mathfrak{H}(Q_{a b}))\bigg)\delta^{b' b}\delta^{a' a}\bigg].
\label{mdb}
\end{eqnarray}	
As Debye mass is color dependent and therefore,  one need to  sum the contributions from all
the colors and then average over the number of colors to get the total Debye mass i.e., 
\begin{equation}
\bar m_D^2=\sum_{abcd} \frac{(m_D^2)_{abcd}}{N^4}.
\label{dbmass1}
\end{equation}
In the large $N$ limit (i.e., neglecting 1/N terms in Eq.(\ref{mdb})), the Debye mass is diagonal and its components can be written in the limit quark mass $m=0$ as
\begin{equation}
(m_D^2)_{1}=(m_D^2)_{3}=\frac{g^2 T^2}{6}(6+N_f-36q+(60-12N_f)q^2),
\end{equation}
\begin{equation}
(m_D^2)_{2}=\frac{g^2 T^2}{6}(N_f+6(1-2q)^2).
\end{equation}
which is same as was derived in Ref.\cite{Hidaka:2009hs}
It is easy to check that, in the limit $ Q=0$ and $m=0$,
the Debye mass as written in Eq.(\ref{mdb}) reduces to its  familiar HTL limit given as
\begin{equation}
(m_D^2)_{abcd}=\frac{g^2 T^2}{3}\bigg(N_c+\frac{N_f}{2}\bigg)\mathcal{P}_{abcd}.
\end{equation}
In our calculation for the heavy quark transport coefficients, however, we will use the color averaged Debye mass as given in Eq.(\ref{dbmass1}).
\subsection{Light Quark thermal mass}
\label{quarkmass}
\begin{figure}[tbh]
\includegraphics[width=12cm]{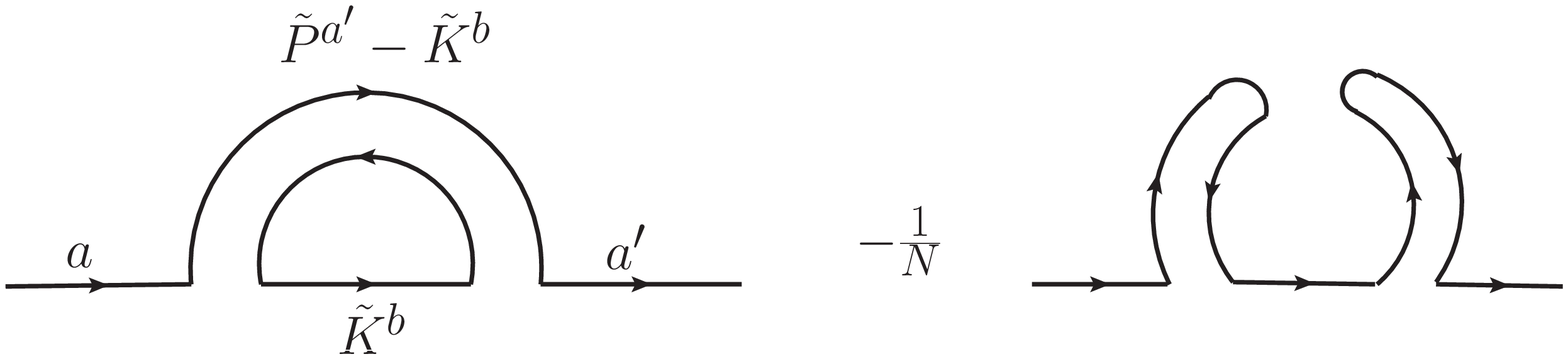}
\caption{One loop quark self energy diagram in double line notation}
\label{selfloop}
\end{figure}

In the double line notation, the standard diagram of one loop quark self energy is shown in Fig.(\ref{selfloop}) where $a$ and $a'$ 
respectively are the color indices for incoming and outgoing quark. It is expected that similar to the gluon self energy, the
 quark self energy also depends on the colors of incoming and outgoing quark and in the presence of a background gauge 
field  the same can be written as
\begin{equation}
\Sigma(P,Q,m)_{a'a}=g^2 (t^{d e})_{a' b} \mathcal{P}_{d e f g} (t^{f g})_{b a} \int \frac{d^4 K}{(2 \pi)^4}\frac{\gamma^{\mu} (m-\slashed{\tilde{K}_b})\gamma_{\mu}} {(\tilde{P}_{a'}-\tilde{K}_b)^2 (\tilde{K}_b^2+m^2)},
\label{quarkself}
\end{equation}
where $g$ is coupling constant, $\tilde{K}_b=K+\tilde{Q}_b$ is quark momentum and $\tilde{P}_{a'}-\tilde{K}_{b}=P-K+\tilde{Q}_a-\tilde{Q}_b$ is gluon momentum. 
To solve the integration in Eq.(\ref{quarkself}), let us first write $\int \frac{d^4K}{(2\pi)^4}=\sum\limits_{n=-\infty}^\infty
 \int\frac{d\zbf k}{(2\pi)^3}; k_4\equiv\omega_n=2n\pi T$ and perform Matsubara frequency sum. There are two 
types of terms where one need to perform frequency summation. One is similar to Eq.(\ref{frq1}) with product of 
two propagators $\sum \Delta(K)\Delta(P-K)$ (arising from the term proportional to $m$) and another is $\sum \omega_n\Delta(K)\Delta(P-K)$ 
(arising from the $\slashed{\tilde{K}}_b$ term). The later one can be written as
\begin{eqnarray}\nonumber 
T\sum_{n}\omega_n\Delta(K)\Delta(P-K)&=&\frac{i}{4 E_q}\bigg(\frac{f(E_q+i Q2+i\omega)-f(E_k-i Q1)}{E_k-E_q-i(Q1+Q2+\omega)}+\frac{1+f(E_k-i Q1)+f(E_q-i Q2-i\omega)}{E_k+E_q-i (Q1+Q2+\omega)}\nonumber\\
&+&\frac{f(E_q-i Q2-i\omega)-f(E_k+i Q1)}{E_k-E_q+i(Q1+Q2+\omega)}
+\frac{1+f(E_q+i Q2+i\omega)+f(E_k+i Q1)}{E_k+E_q+i(Q1+Q2+\omega)}\bigg).
\label{fsum1} 
\end{eqnarray}
We take HTL approximation and evaluate only $T^2$ dependent term in quark-self energy. We note here that, unlike gluon self energy, one does not get any extra term different in structure as compared to the usual perturbative HTL approximation for the quark self energy. The leading contribution arises from the terms having $E_q-E_k$ in the denominators  of Matsubara frequency sums and in Eq.(\ref{fsum1}) comes from the first and the third terms.  Simplifying Eq.(\ref{quarkself}) with Eqs.(\ref{frq1}) and (\ref{fsum1}), quark self energy becomes 
\begin{eqnarray}
\Sigma(P,Q,m)_{a'a}&=&g^2 \mathcal{P}_{a' b, b a} \bigg(m \int\frac{d \zbf{k}}{(2 \pi)^3}\frac{1}{4 E_k E_q}\bigg[\frac{f(E_q-i Q2)+f(E_q+i Q2)}{P_{a}.\hat{K}}\nonumber\\
&-&\frac{f(E_k+i Q1)+f(E_k-i Q1)}{P_{a}.\hat{K}}\bigg]+\int\frac{\slashed{\hat{K}}d^3k}{E_k(2 \pi)^3}\bigg[\frac{f(E_k+i Q2)-f(E_q-i (Q1+\omega))}{P_{a}.\hat{K}}\nonumber\\
&-&\frac{f(E_q+i (Q1+\omega))-f(E_k-i Q2)}{P_{a}.\hat{K}}\bigg]\bigg).
\label{quarkself1}
\end{eqnarray}
In the above equation, we have used HTL approximation so that $E_q-E_k \approx-\frac{\vec{P}.\vec{k}}{E_k}$, $f(E_k-i Q) \approx f(E_q-i Q) $ 
and $e^{\frac{i\omega}{T}}\simeq 1$. Here $Q1=\tilde{Q}_b$, $Q2=Q_{a'}-Q_b$ and $\hat{K}=(i,\hat{k})$. After simplifying Eq.(\ref{quarkself1})  further, it can be written as
\begin{equation}
\Sigma(P,Q,m)_{a'a}=\frac{g^2 T^2}{8 \pi^2} \sum_{b=1}^{3}\mathcal{P}_{a' b, b a} 
\bigg([\mathfrak{F}(q2,y)-\mathfrak{F}({q}1,y)]\int\frac{d\Omega}{4\pi}\frac{\slashed{\hat{K}}}{P_{a}.\hat{K}}+\frac{m}{T}
(\mathfrak{J}(q2,y)-\mathfrak{J}({q}1,y))\int\frac{d\Omega}{4 \pi}\frac{1}{P_{a}.\hat{K}}\bigg),
\label{sigeq}
\end{equation}
where as before, $y=\beta m$, $q1=\beta Q1$; $\mathfrak{F}(q)$ is same as given in Eq.(\ref{funfF}) and $\mathfrak{J}$ is given as
\begin{equation}
\mathfrak{J}(q,y)=\int\frac{x^2 dx}{{x^2+y^2}}(f(x,y,-iq)+f(x,y,iq)).
\end{equation} 
It is easy to see that to estimate the quark thermal mass from its self energy, one need to sum over colors in Eq.(\ref{sigeq}) keeping $a$ and $a'$ open indices. After performing this color sum, quark self energy reduces to
\begin{eqnarray}
\Sigma(P,Q,m)_{a'a}&=&\frac{g^2 T^2}{8 \pi^2} \delta_{a' a}\bigg(\bigg[\sum_{b=1}^{3}(\mathfrak{F}(q_{a' b},y)-\mathfrak{F}(\tilde{q}_b,y))
-\frac{1}{3}(\mathfrak{F}(0,y)-\mathfrak{F}(\tilde{q}_a,y))\bigg]\int\frac{d\Omega}{4 \pi}\frac{\slashed{\hat{K}}}{P_{a}.\hat{K}}\nonumber\\
&+&\frac{m}{T} \bigg[\sum_{b=1}^{3}(\mathfrak{J}(q_{a' b},y)-\mathfrak{J}(\tilde{q}_b,y))+ \mathfrak{J}(0,y)-\mathfrak{J}(\tilde{q}_a,y)\bigg]\int\frac{d\Omega}{4 \pi}\frac{1}{P_{a}.\hat{K}}\bigg).
\label{qmass}
\end{eqnarray}
In the HTL approximation, the effective fermion mass (thermal mass) can be written as \cite{Thoma:2000dc}
\begin{equation}
4 m_{th}^2= Tr(\slashed{P} \Sigma(P)).
\label{qmass1}
\end{equation}
From Eqs.(\ref{qmass1}) and (\ref{qmass}), the color dependent quark thermal mass a function of Polyakov loop parameter $q$ can be written as
\begin{eqnarray}
m_{a'}^2&=&\frac{g^2 T^2}{8 \pi^2} \bigg(\sum_{b=1}^{3}(\mathfrak{F}(Q_{a' b},y)-\mathfrak{F}(\tilde{Q}_b,y))-\frac{1}{3}(\mathfrak{F}(0,y)-\mathfrak{F}(\tilde{Q}_{a'},y))\bigg).
\label{qthm}
\end{eqnarray}
In the limit of vanishing quark mass, using Eq.(\ref{polylog}), it is easy to show that
\begin{equation}
m_{a}^2=\frac{g^2 T^2}{6}\bigg(1+\frac{3}{2}q_{a}+\frac{7}{2}q_{a}^2\bigg).
\end{equation}
In the subsequent calculationS that follow, we however, keep the quark mass dependence as in Eq.(\ref{qthm}). Similar to Eq.(\ref{dbmass1}), one can define a color averaged quark thermal mass as
\begin{equation}
m_{th}^2=\sum_{a=1}^{3} \frac{m_a^2}{3}
\label{mth}
\end{equation}
so the total quark mass becomes
\begin{equation}
m_q=m+m_{th},
\label{mquark}
\end{equation}
Thus the color averaged Debye mass for the gluons and color averaged thermal mass for quarks as given by Eqs.(\ref{dbmass1}) 
and (\ref{mth}) depend upon the Polyakov loop parameter.

\begin{figure}[tbh]
\subfigure{
\includegraphics[width=7cm,height=6cm]{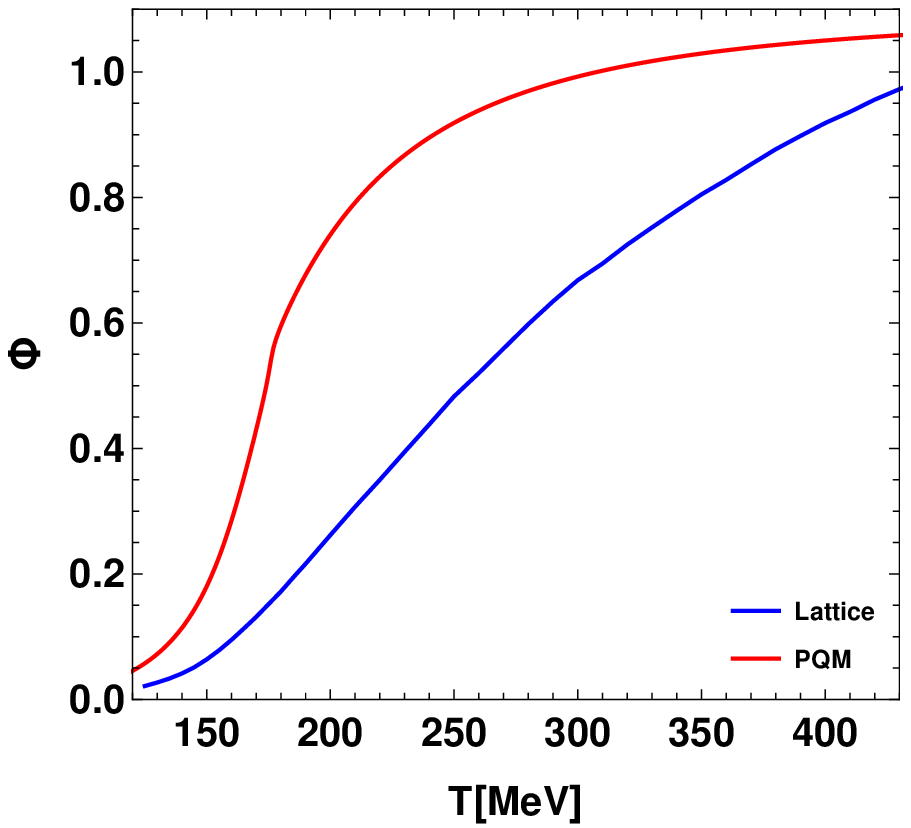}}
\subfigure{
\includegraphics[width=7.16cm,height=6.3cm]{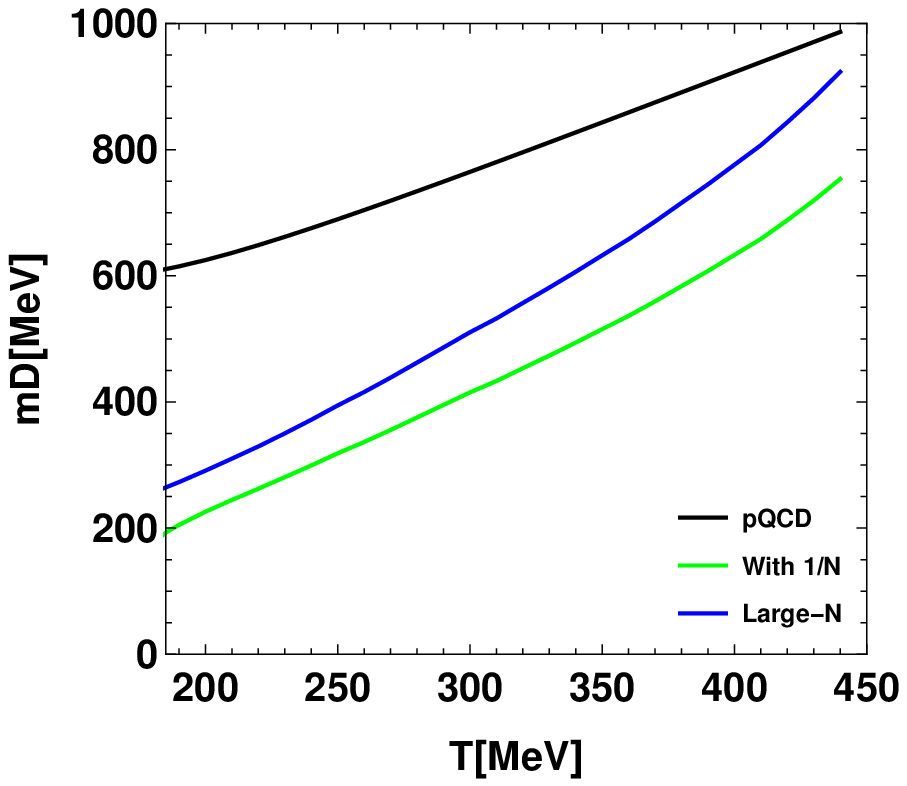}}
\caption{\textbf{Left panel:} Polyakov loop value as a function of temperature. The red curve is from PQM model~\cite{Abhishek:2017pkp}. The blue curve is from the lattice results of Ref.\cite{Bazavov:2016uvm}. \textbf{Right panel:} Debye mass ($m_{D}$) as a function of temperature. The black curve corresponds to pQCD hard thermal loop calculations~\cite{Bellac:2011kqa}. The blue curve corresponds to large $N$ limit for $m_{D}$ as given in Eq.(\ref{mdb}). The green curve correspond to taking all the terms in Eq.(\ref{mdb}) for $N=3$. Here Polyakov loop is taken from lattice data~\cite{Bazavov:2016uvm}.}
\label{ploop}
\end{figure}

For the Polyakov loop parameter, we adopt here two approaches. Firstly, we estimate the same from a phenomenological 2 flavor PQM 
model ~\cite{Abhishek:2017pkp,bjschaefer}. The salient features of the model and the parameters taken in the model is discussed 
in Appendix A. With this parameterization the critical temperature for the crossover transition $T_c\approx 176$ MeV. 
We also take the Polyakov loop parameter from lattice simulations as in Ref.~\cite{Bazavov:2016uvm}. The variation of the 
Polyakov loop with temperature (T) is shown on the left of Fig(\ref{ploop}). 
Clearly, compared to the lattice simulations, the Polyakov loop parameter $\phi$ in PQM model shows a sharper rise and
reaches its asymptotic value $\phi=1$ at the temperature around 320 MeV. On the other hand, in the lattice simulations, 
this happens at a much higher temperature. This means that the non-perturbative effects are significant up to temperature 
as high as $~400 $ MeV in lattice. However in PQM these effects are significant only temperatures upto around $~320 $ MeV. 
On the right side of Fig(\ref{ploop}), Debye mass as a function of temperature is shown. Here the black curve corresponds to the 
Debye mass in pQCD, while the blue and the green curves are in the presence of Polyakov loop. The blue curve
corresponds to the large N limit (i.e., dropping 1/N terms in Eq.(\ref{selffer1})). On the other hand, the Green curve
corresponds to including  the 1/N terms in Eq.(\ref{selffer1}). Clearly, the large N limit approaches the perturbative limit 
faster compared to the one including $1/N$ terms for the Debye mass.  
 
\begin{figure}[tbh]
\subfigure{
\includegraphics[width=7.16cm]{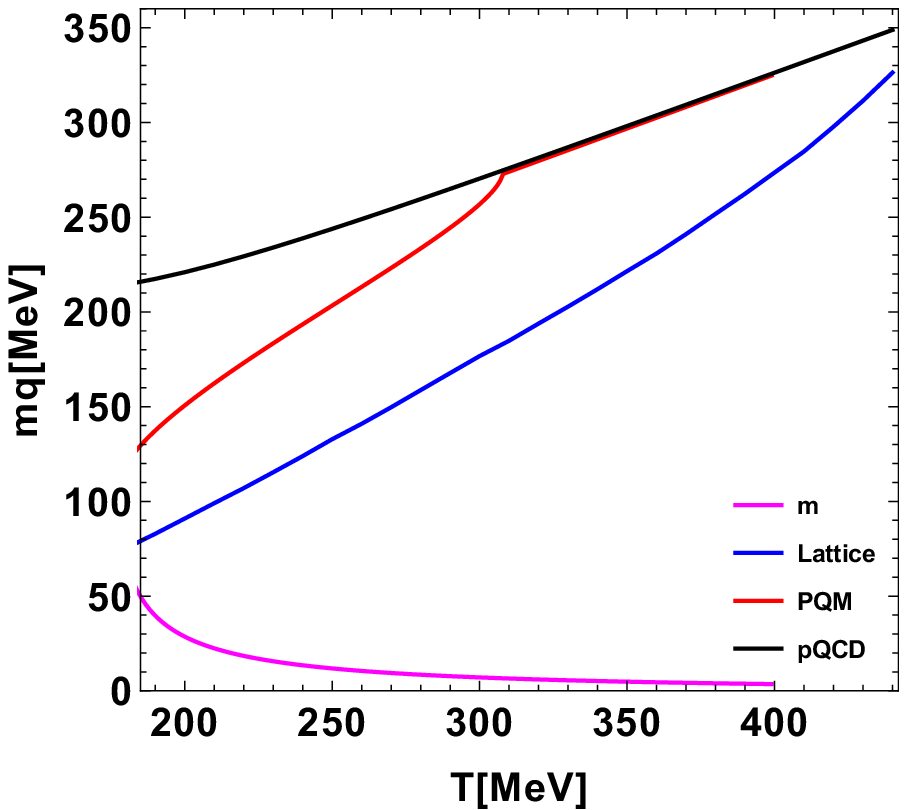}}
\subfigure{
\hspace{-0mm}\includegraphics[width=7.33cm]{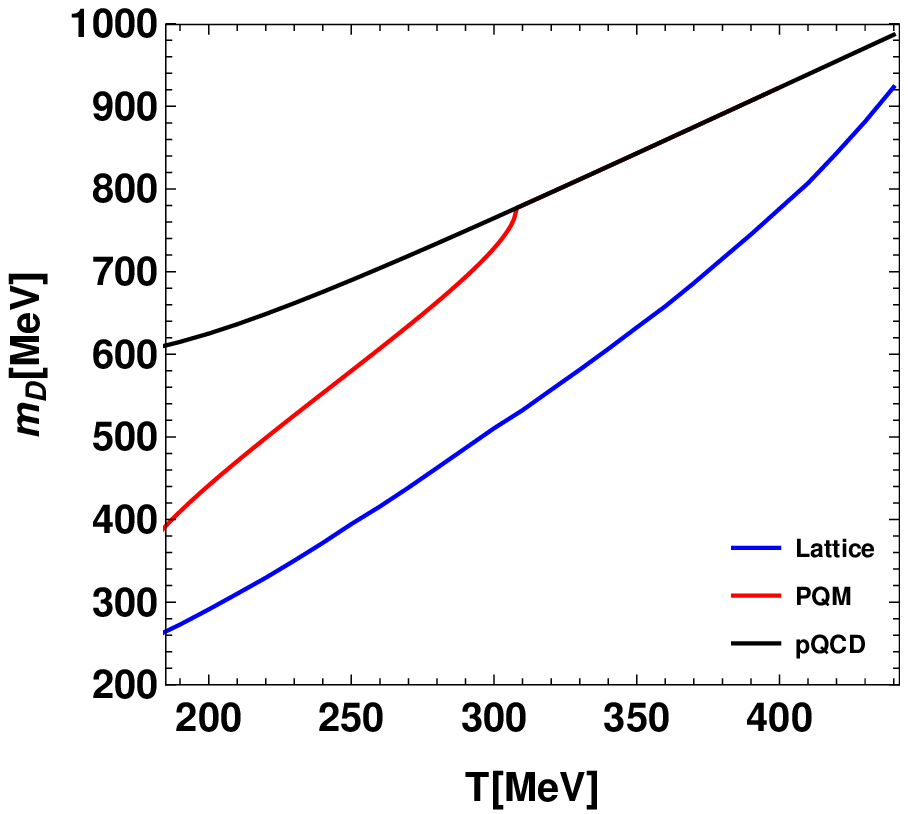}} 
\caption{
\textbf{Left panel:} Quark masses as a function of temperature. The bottom-most curve (magenta) shows the constituent quark mass estimated in PQM model. The topmost curve (black) shows the perturbative HTL estimate of quark thermal mass~\cite{Bellac:2011kqa}. The red curve shows the temperature dependence of quark thermal mass (Eq.(\ref{mth})) with the Polyakov loop taken from PQM model calculations. The blue curve shows the thermal mass of quark (Eq.(\ref{mth})) using Polyakov loop from lattice simulations~\cite{Bazavov:2016uvm}. \textbf{Right panel:} Debye mass as a function of temperature in the leading order in N of Eq.(\ref{mdb}). The blue curve correspond to Polyakov loop value taken from lattice data~\cite{Bazavov:2016uvm} while the red curve correspond to the Polyakov loop value taken from PQM model.}
\label{mD}
\end{figure}

For the light quarks, different contributions to the masses as a  function of temperature ($T$) are shown on the left side of Fig. (\ref{mD}). Red and blue curves correspond to quark thermal masses ($m_{th}$) as given in Eq.(\ref{mth}) evaluated in the HTL 
approximation in the presence of a background gauge field. The red curve corresponds to Polyakov loop value taken from PQM 
model while the blue curve corresponds to the same taken from lattice simulations. The HTL perturbative QCD thermal
masse as in Ref.~\cite{Bellac:2011kqa} is shown by the black curve. Clearly, with the lattice value of the Polyakov loop, thermal masses 
approach the perturbative results at a much higher temperature while with values taken from PQM, the perturbative limit 
reaches at a relatively lower temperature around 320 MeV. It ought to be mentioned that beyond 330 MeV $\phi$ value 
is larger than one in which case $q$ becomes imaginary. We have taken here the real part of $q$ for estimating the thermal masses. 
Beyond temperature 330 MeV the real part of $q$ vanishes which leads to the perturbative limit. As compared to PQM model, 
the color averaged thermal mass is smaller for Polyakov loop expectation value taken from lattice simulation. This is because, 
with the smaller value of $\phi$, statistical distribution functions are suppressed more.  The magenta curve is the  constituent 
quark mass estimated in PQM model. The right side of Fig.(\ref{mD}), shows the behavior of color averaged Debye mass in the large N limit of Eq(\ref{mdb}). The red and the blue curves correspond to the masses with Polyakov loop value taken from PQM and lattice simulations respectively. Debye mass is smaller as compared to the perturbative QCD Debye mass and this suppression is more when $\phi$ is taken from the lattice simulations. The reason for this is the same as that for the case of quark thermal mass. In the estimation of the transport coefficients, we shall use the Debye mass and thermal masses of quarks as in Eq.(\ref{mquark}). It is clear that the non-perturbative effects which are in the distribution function and
the masses of quarks and gluons can significantly affect these transport coefficient as compared to the perturbative QCD.

\section{Results and discussions}
\label{results}
With the thermal mass of the quarks and the Debye mass as computed in the background of a nontrivial Polyakov loop, we next numerically compute the drag and diffusion coefficients using Eq.(\ref{transport111}).
For the heavy quark elastic interaction with the light quarks and gluons, $qQ\rightarrow qQ$ and $gQ \rightarrow gQ$ 
scattering processes are considered where $Q$ stands for heavy quark, $q$ stands for light quarks and $g$ stands for the gluon. 
In the case of massless light quark and gluon, the leading-order (LO) matrix elements for $qQ\rightarrow qQ$ and 
$gQ \rightarrow gQ$ scattering have been calculated in Ref.~\cite{Combridge, Svetitsky:1987gq}. These pQCD 
cross sections have to be supplemented by the value of the coupling constant and the Debye screening mass which is 
needed to shield the divergence associated with the $t$-channel diagrams to compute the heavy quark transport coefficients. 
For massive light quark and gluon, the calculation of the scattering matrix, ${\cal M}_{{(q,g)+Q} \rightarrow {(q,g)+Q}}$,  is  performed considering the leading-order (LO) diagram with massive quark and gluon propagators for $gQ \rightarrow gQ$ and a massive gluon propagator for $qQ\rightarrow qQ$ scatterings~\cite{Berrehrah:2013mua, Scardina:2017ipo}.  Within the matrix model, the scattering amplitudes are summarised in Appendix(B). Similar to previous work~\cite{Berrehrah:2013mua, Scardina:2017ipo}, massive gluon propagator for $qQ\rightarrow qQ$ and t-channel of $gQ\rightarrow gQ$ is used. We estimate the transport coefficients for the charm 
quark whose mass is taken as $m_C=1.27$ GeV.
Here we use the two loop running coupling constant given as ~\cite{{Caswell:1974gg}}
\begin{equation}
\alpha_s=\frac{1}{4\pi}\bigg(\frac{1}{2 \beta_{0} \ln(\frac{\pi T}{\Lambda})+\frac{\beta_1}{\beta_0}\ln(2 \ln(\frac{\pi T}{\Lambda}))}\bigg)
\end{equation}
where 
\begin{equation}
\beta_0= \frac{1}{16 \pi^2}\bigg(11-\frac{2 N_f}{3}\bigg)
\end{equation}
\begin{equation}
\beta_1=\frac{1}{(16 \pi^2)^2}\bigg(102-\frac{38 N_f}{3}\bigg)
\end{equation}
with $\Lambda=260$ MeV and $N_f=2$.

\begin{figure}[tbh]
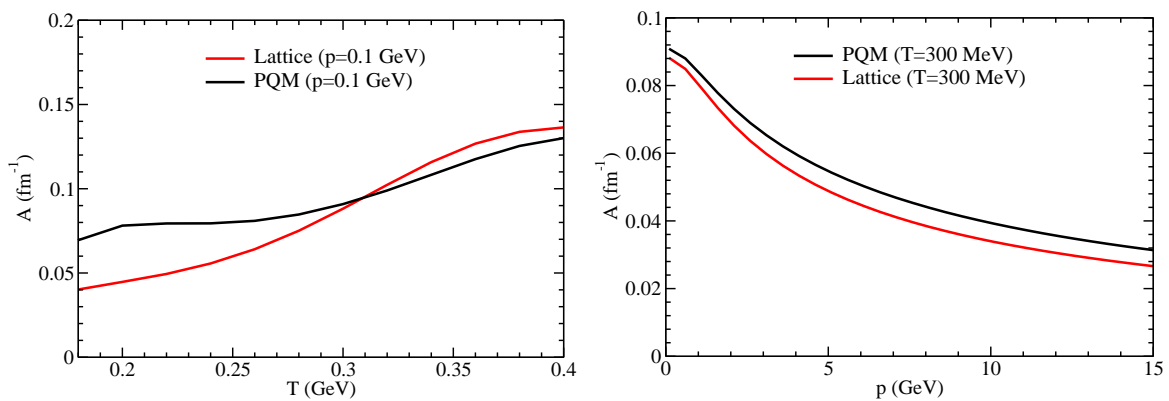

\subfigure{
\includegraphics[width=7.5cm]{Drag_PL.eps}}
\subfigure{
\hspace{-0mm}\includegraphics[width=7.5cm]{Drag_PL_p.eps}} 
\caption{Variation of drag coefficients (A) with temperature (left) for momentum $p=100$ MeV and with momentum (right) for temperature $T=300$ MeV.}
\label{dragcomp}
\end{figure}

We evaluate the drag and diffusion coefficients of heavy quark in QGP with Polyakov loop value from two different models. 
In one case the Polyakov loop value, hence the Debye mass and thermal masses,  has been taken from PQM calculation as inputs to compute the heavy quark transport and  we label it as PQM. In the other case, 
Polyakov loop value has been taken from the lattice simulatons and hence, we label it as lattice in the following discussions. 
The temperature variation of the drag coefficient has been shown in Fig.(\ref{dragcomp}) for charm quark interaction with light quarks and gluon for a given momentum (p=0.1 GeV) obtained for both PQM and lattice Polyakov loop values.

\vspace{1cm}

\begin{figure}[tbh]
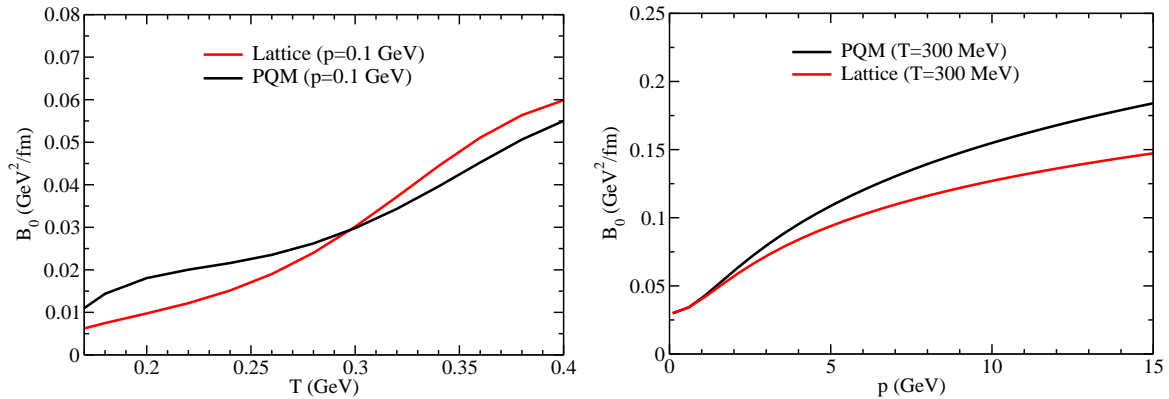

\subfigure{
\includegraphics[width=7.5cm]{Diffc_PL.eps}}
\subfigure{
\hspace{-0mm}\includegraphics[width=7.5cm]{Diffc_PL_p.eps}} 
\caption{Variation of diffusion coefficients ($B_0$) with temperature (left) for momentum $p=100$ MeV and with momentum (right) for temperature $T=300$ MeV.}
\label{diffcomp}
\end{figure}

We obtain quite a mild temperature dependence of  
heavy quark drag coefficient for the case of PQM. However, with lattice, we obtained a quite stronger 
temperature dependence of heavy quark drag coefficient than the one with PQM. We notice that the drag coefficient 
obtained with PQM input is larger at low temperature than the one obtained with lattice inputs whereas the trend 
is opposite at high temperature. This is mainly because of the interplay between the Debye mass and Polyakov loop 
value obtained within both the models.

A smaller value of the Polyakov loop, as shown on the left side of Fig.(\ref{ploop}),  in case of lattice reduces the magnitude of the drag coefficients at low temperature. However, at high temperature, with smaller Debye mass, as shown in Fig.(\ref{mD}), obtained with lattice input enhances the magnitude of heavy quark drag coefficients. Hence, at low temperature Polyakov loop value plays the dominant role ( e.g., at T=180 MeV, the Polyakov loop value obtained within both the models differ by a factor about 2) whereas at high temperature the Debye mass plays the dominant role ( e.g., T=300 MeV, the differences between the Polyakov loop value obtained within both cases reduced significantly) for the behavior of the drag coefficient.

We observed temperature dependence of  heavy quark drag coefficient obtained with PQM Polyakov loop value is quite consistent with the results obtained with other quasi-particle models ~\cite{Das:2015ana, Berrehrah:2013mua} and T-matrix approach ~\cite{vanHees:2007me}. It is important to mention that the temperature dependence of the drag coefficient plays a significant role ~\cite{Das:2015ana} to describe heavy quark $R_{AA}$ and $v_2$ simultaneously, which is a challenge to almost all the models on heavy quark dynamics. A constant or weak temperature dependence of the drag coefficient is an essential ingredient to reproduce the heavy quarks $R_{AA}$ and $v_2$ simultaneously, whereas in pQCD the drag coefficient increases with temperature.

The momentum variation of the drag coefficient has been shown in the right panel of Fig.(\ref{dragcomp}) for charm quark interaction with light quarks and gluon obtained with PQM and lattice Polyakov loop value. We observe a strong momentum dependence of heavy quark drag coefficient as compared to the same estimated within pQCD~\cite{Rapp:2018qla}. This is mainly due to the inclusion of non-perturbative effects through the Polyakov loop background. At T=300 MeV the drag obtained with the PQM Polyakov loop (at p=0.1 GeV) is marginally larger than the drag obtained with lattice Polyakov value. Hence, the momentum variation of drag coefficients obtained with inputs from PQM  is marginally larger than the one obtained with inputs from lattice simulation in the entire momentum range considered here.

In Fig.(\ref{diffcomp}) heavy quark diffusion coefficient $B_0$ has  been displayed  as a function of temperature obtained with input parameter from PQM and lattice. The diffusion coefficients increases with temperature for both the cases as it involves the square of the momentum transfer. In terms of magnitude the diffusion coefficient obtained within both the cases follow similar trend of drag coefficient due to the same reason (i.e., interplay between Debye mass and Polyakov loop value). 

\vspace{1cm}
\begin{figure}[tbh]
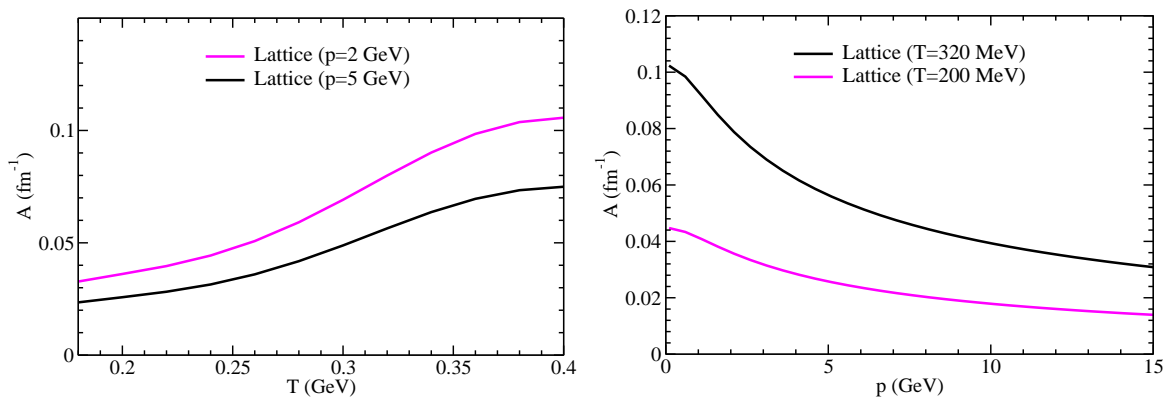

\subfigure{
\includegraphics[width=7.5cm]{Drag_PL_lattice.eps}}
\subfigure{
\hspace{-0mm}\includegraphics[width=7.5cm]{Drag_PL_p_lattice.eps}} 
\caption{Variation of drag coefficients (A) with temperature (left) for different values of momentum  and with momentum (right) for different values of temperature.  The Polyakov loop value is taken from the lattice data~\cite{Bazavov:2016uvm}.}
\label{draglatt}
\end{figure}

\begin{figure}[tbh]
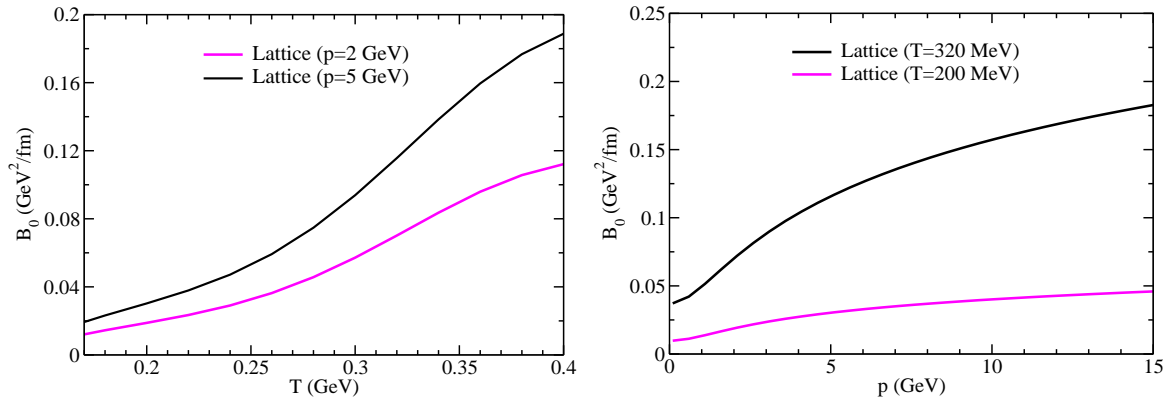

\subfigure{
\includegraphics[width=7.5cm]{Diffc_PL_lattice.eps}}
\subfigure{
\hspace{-0mm}\includegraphics[width=7.5cm]{Diffc_PL_p_lattice.eps}}\caption{Variation of diffusion coefficients ($B_0$) with temperature for different values of the momentum(left) and with momentum (right) for different values of temperature. The Polyakov loop value is taken from the lattice data~\cite{Bazavov:2016uvm}.}
\label{difflatt}
\end{figure}

The momentum variation of the diffusion coefficient has been shown in  Fig.(\ref{diffcomp}) for charm quark interaction with light quarks and gluons for the same values of Polyakov loop.  Similar to the drag coefficient, the diffusion coefficient also shows the same trend with PQM having larger value then that from the lattice as a function of momentum. Stronger suppression of distribution function at high momentum in lattice Polyakov loop than that of from PQM also play a marginal role in the momentum variation of heavy quark drag and diffusion coefficients obtained. 

To understand the temperature dependence of the transport coefficients, 
we plot the temperature variation of the drag coefficient in Fig.~\ref{draglatt} at different momentum 
obtained with Polyakov loop value from lattice simulations. We obtain almost similar temperature dependence of heavy quark drag coefficient 
at both the momentum having larger magnitude at p=2 GeV than at p=5 GeV. In Fig.~\ref{difflatt} we have depicted 
the temperature variation of diffusion coefficient at different momentum for the same values of Polyakov loop.
As expected, the magnitude of the diffusion coefficient is large at p=5 GeV than p=2 GeV having similar 
temperature variation for both the momenta.

In Fig.~\ref{draglatt} we have shown the variation of drag coefficient with momentum at different temperature 
obtained with the lattice inputs. We observe a larger magnitude of the drag coefficient at T=320 MeV than T=200 MeV 
but the momentum variation is similar at both temperature. Momentum variation of the diffusion coefficient has 
been depicted in Fig.~\ref{difflatt} at difference temperature. At both the momenta the 
diffusion increase with temperature having larger magnitude at T=320 MeV than T=200 MeV.

It is worth  mentioning here that, non-perturbative effects from a  different perspective 
has been investigated recently in Ref.~\cite{Liu:2016ysz, Liu:2017qah, Liu:2018syc} and employed to calculate the 
transport coefficients~\cite{Liu:2018syc}. The method here consisted of using T-matrix with an in-medium potential for the 
heavy quarks. This potential is constrained by the heavy quark free energy from the lattice data. The lattice heavy quark free energy
 is directly related to the Polyakov loop and hence is correlated with the strength of the confining potential. Therefore it is
 nice to see that the behavior of drag coefficient being rather flat with regards to temperature dependence whereas the
 diffusion coefficient having a strong temperature dependence as observed here was also observed in Ref.\cite{Liu:2018syc}.
 This consistency suggest of having a possible existence of model independent correlation between Polyakov loop and the heavy
 quark transport coefficients.

\section{Summary} 
\label{summary}
In this work, we have computed the heavy quark drag and diffusion coefficients in QGP including non-perturbative effects
via a Polyakov loop background. In order to incorporate these effects we first calculate quark and gluon thermal masses also taking the quark constituent mass into account. We found that for temperatures below 300 MeV quark thermal mass and gluon Debye mass
starts deviating from its perturbative value this effect significant for even higher temperatures when Polyakov values are
taken from the lattice simulations. This decrease in the Debye mass of gluon and the thermal mass of light quarks is due to 
color suppression manifested in the quark and gluon distribution functions in the presence of a background Polyakov loop field.
In the calculation of HQ diffusion coefficient the distribution function of the light quark and the Debye mass play
complimentary roles. While the distribution function with Polyakov loop tend to decrease the HQ transport coefficient
the Debye mass has the effect of increasing these transport coefficients.  We have found a weak temperature dependence of the 
heavy quark drag coefficient with Polyakov loop value taken from PQM  which is consistent with other models like T-matrix and
quasi particle model which also take into account the non-perturbative effects in a different manner. 
This consistency suggests existence of possible model independent correlations between the results obtained with
the Polyakov loop and other non-perturbative models and reaffirm the temperature and momentum dependence of heavy
quark transport coefficients. In the present investigation, we have 
aconfined our attention to the elastic $2\rightarrow 2$ processes within the matrix model. Inclusion of other effects arising from 
$2\rightarrow 3$ processes, LPM effects are expected to be sub-dominant due to the large mass of the heavy quark\cite{Zhang:2003wk}
but, none the less, can be important 
at high parton density.
We plan to explore different possible phenomenological implications of the present investigation in future.

\vspace{1cm}

{\bf Acknowledgment:} 
S.K.D. acknowledges the support by the National Science Foundation of China (Grants No. 11805087 and No. 11875153). We would like to thank Yoshimasa Hidaka for valuable discussions and important clarifications on the HTL resummed propagator in presence of Polyakov loop.

\appendix
\section{Polyakov loop extended Quark Meson model}
\label{PQM}
Polyakov loop extended quark meson model(PQM) captures two important features of quantum chromodynamics(QCD) - namely chiral symmetry 
breaking and its restoration at high temperature and/densities 
as well as the confinement - deconfinement transitions. 
Explicitly, the Lagrangian of the PQM model is given by\cite{bjschaefer,guptatiwari,bielich,buballa,ranjita}
\bearr
{\cal L}&=&\bar\psi\left(i\gamma^\mu D_\mu-m-g_\sigma(\sigma+i\gamma_5\bfm\tau\cdot\bfm\pi)\right)\psi\nonumber 
+ \frac{1}{2}\left[\partial_\mu\sigma\partial^\mu\sigma+\partial_\mu\bfm\pi\partial^\mu\bfm\pi\right]\nonumber - U_\chi(\sigma,\bfm\pi)-U_P(\phi,\bar\phi)\\
\label{lagpqm}
\eearr
In the above, the first term is the kinetic and interaction term for the quark doublet $\psi=(u,d)$ interacting with the scalar ($\sigma$)
and the isovector pseudoscalar pion $({\bfm \pi})$ field. The scalar field $\sigma$ and the pion field ${\bfm\pi}$ together form a SU(2)  isovector field.
The quark field is also coupled to a spatially constant temporal gauge field $A_0$ 
through the covariant derivative $D_\mu=\partial_\mu-ieA_\mu$; $A_\mu=\delta_{\mu 0}A_\mu$.

The mesonic potential $U_\chi(\sigma,\zbf \pi)$ essentially describes 
the chiral symmetry breaking pattern in strong interaction and is given by
\be
U_\chi(\sigma,\bfm \pi)=\frac{\lambda}{4}(\sigma^2+\bfm \pi^2-v^2)-c\sigma
\label{uchi}
\ee

The last term in the Lagrangian in Eq.(\ref{lagpqm}) is responsible for including the physics of color confinement
in terms of a potential energy for the expectation value of the 
Polyakov loop $\phi$ and $\bar\phi$ which are defined in terms 
of the Polyakov loop operator which is a Wilson loop in the temporal direction
\be
{\cal P}=P\exp\left( i\int_0^\beta dx_0 A_0(x_0,\zbf x)\right).
\ee
In the Polyakov gauge  $A_0$ is time independent and is in the Cartan subalgebra i.e. $A_0^a=A_0^3\lambda_3+A_0^8\lambda_8$.
One can perform the integration over the time variable trivially as path ordering 
becomes irrelevant so that ${\cal P}(\zbf x)=\exp(\beta A_0)$.
The Polyakov loop variable $\phi$ and its hermitian conjugate $\bar\phi$ 
are defined as
\be
\phi(\zbf x)=\frac{1}{N_c}Tr {\cal P(\zbf x)\quad\quad \bar\phi(\zbf x)}=\frac{1}{N_c}{\cal P}^\dagger(\zbf x).
\ee
In the limit of heavy quark mass, the confining phase is center symmetric and therefore $\langle\phi\rangle=0$ while for deconfined phase 
$\langle \phi\rangle\neq 0$. Finite quark masses break this symmetry explicitly.
The explicit form of the potential $U_p(\phi,\bar\phi)$ is not known 
from first principle calculations. The common strategy is to choose a functional form of the potential that reproduces the pure gauge lattice
simulation thermodynamic results. Several forms of this potential has been suggested in literature. We shall use here the following
polynomial parameterization \cite{bjschaefer}
\be
U_P(\phi,\bar\phi)=T^4\left[-\frac{b_2(T)}{2}\bar\phi\phi-\frac{b_3}{2}(\phi^3+\bar\phi^3)+\frac{b_4}{4}(\bar\phi\phi)^2\right]
\label{uphi}
\ee
with the temperature dependent coefficient $b_2$ given as
\be
b_2(T)=a_0+a_1(\frac{T_0}{T})+a_2(\frac{T_0}{T})^2+a_3(\frac{T_0}{T})^3
\ee
The numerical values of the parameters are
\bearr
&&a_0=6.75,\quad a_1=-1.95, \quad a_2=2.625, \quad a_3=-7.44\nonumber\\
&&b_3=0.75, \quad b_4=7.5
\\
\label{parameters}
\eearr
The parameter $T_0$ corresponds to the transition temperature of Yang-Mills theory. However, for the full
dynamical QCD, there is a flavor dependence on $T_0(N_f)$. For two flavors we take it to be $T_0(2)=192$ MeV as in
Ref.\cite{bjschaefer}.

The Lagrangian in Eq.(\ref{lagpqm}) is invariant under  $SU(2)_L\times SU(2)_R$ transformation when the
explicit symmetry breaking term $c\sigma$ vanishes in the potential $U_\chi$ in Eq.(\ref{uchi}).
The parameters of the potential $U_\chi$ are chosen such that the chiral symmetry is spontaneously broken in the vacuum.
The expectation values of the meson fields in vacuum are $\langle\sigma\rangle=f_\pi$ and $\langle\bfm\pi\rangle=0$. Here $f_\pi=93$ MeV 
is the pion decay constant. The coefficient of the symmetry breaking linear term is decided from the partial conservation of
axial vector current (PCAC) as 
$c=f_\pi m_\pi^2$, $m_\pi=138$ MeV, being the pion mass. Then minimizing the potential one has $v^2=f_\pi^2-m_\pi^2/\lambda$.
The quartic coupling for the meson,
$\lambda$ is determined from the mass of the sigma meson given as $m_\sigma^2=m_\pi^2+2\lambda f_\pi^2$. In the present work we take $m_\sigma=600 $MeV which gives $\lambda$=19.7. The coupling $g_\sigma$ is fixed here from the constituent quark mass in vacuum $M_q=g_q f_\pi$
which has to be about (1/3)rd of nucleon mass that leads to $g_\sigma=3.3$ \cite{rischkepqm}.

To calculate the bulk thermodynamical properties of the system we use 
a mean field approximation for the meson and the Polyakov fields while retaining the quantum and thermal fluctuations of the quark fields. The thermodynamic potential can then be written as
\begin{equation}
\Omega(T,\mu)=\Omega_{\bar q q}+U_\chi+U_P(\phi,\bar\phi)
\label{thpot}
\end{equation}
The fermionic part of the thermodynamic potential is given as
\begin{equation}
\Omega_{\bar q q}=-2N_fT\int \frac{d^3 p}{(2 \pi)^3} 
\left[\ln\left(1+3(\phi+\bar\phi e^{-\beta\omega_-})e^{-\beta\omega_-}+e^{-3\beta\omega_-}\right)\\
+\ln\left(1+3(\phi+\bar\phi e^{-\beta\omega_+})e^{-\beta\omega_+}+e^{-3\beta\omega_+}\right)\right]
\end{equation}
modulo a divergent vacuum part. In the above, $\omega_\mp=E_p\mp\mu$, with the single particle quark/anti-quark energy $E_p=\sqrt{\zbf p^2+M^2}$.
The constituent quark/anti-quark mass is defined to be
\begin{equation}
M^2=g_\sigma^2(\sigma^2+\zbf \pi^2).
\end{equation}
The divergent vacuum part arises from the negative energy states of the Dirac sea. Using standard renormalisation,
it can be partly absorbed in the coupling $\lambda$ and $v^2$. However, a logarithmic correction from the renormalisation
scale remains which we neglect in the calculations that follow \cite{rischkepqm}. 

The mean fields are obtained by minimizing $\Omega$ with respect to $\sigma$, $\phi$, $\bar\phi$, and $\pi$. Extremising the effective potential with respect to $\sigma$ field leads to
\begin{equation}
\lambda(\sigma^2+\zbf\pi^2-v^2)-c+g_\sigma\rho_s=0
\label{gapsigma}
\end{equation}
where, the scalar density $\rho_s=-\langle\bar\psi\psi\rangle$ is given by
\begin{equation}
\rho_s=6N_fg_\sigma\sigma\int\frac{d\zbf p}{(2\pi)^3}\frac{1}{E_P}\left [f_-(\zbf p)+f_+(\zbf p)\right].
\label{rhos}
\end{equation}
In the above, $f_\mp(\zbf p)$ are the distribution functions for the quarks and anti quarks given as
\begin{equation}
f_-(\zbf p)=\frac{\phi e^{-\beta\omega_-}+2\bar\phi e^{-2\beta\omega_-}+ e^{-3\beta\omega_-}}
{1+3\phi e^{-\beta\omega_-}+3\bar\phi e^{-2\beta\omega_-} + e^{-3\beta\omega_-}},
\end{equation}
and,
\begin{equation}
f_+(\zbf p)=\frac{\bar\phi e^{-\beta\omega_+}+2\phi e^{-2\beta\omega_+}+ e^{-3\beta\omega_+}}{1+3\bar\phi e^{-\beta\omega_+}+3\phi e^{-2\beta\omega_+} + e^{-3\beta\omega_+}},
\end{equation}

The condition  $\frac{\partial\Omega}{\partial\phi}=0$ leads to
\begin{equation}
T^4\left[-\frac{b_2}{2}\bar\phi-\frac{b_3}{2}\phi^2+\frac{b_4}{2}\bar\phi\phi\bar\phi\right]+I_\phi=0
\label{gapphi}
\end{equation}
where ,
\begin{equation}
I_\phi=\frac{\partial\Omega_{\bar q q}}{\partial \phi}=-6N_fT\int\frac{d\zbf p}{(2\pi)^3}
\left [\frac{e^{-\beta\omega_-}}{1+3\phi e^{-\beta\omega_-}+3\bar\phi e^{-2\beta\omega_-} + e^{-3\beta\omega_-}}
+\frac{e^{-2\beta\omega_+}}{1+3\bar\phi e^{-\beta\omega_+}+3\phi e^{-2\beta\omega_+} + e^{-3\beta\omega_+}}\right]
,
\end{equation}
Similarly, $\frac{\partial\Omega}{\partial\bar\phi}=0$ leads to
\begin{equation}
T^4\left[-\frac{b_2}{2}\phi-\frac{b_3}{2}\bar\phi^2+\frac{b_4}{2}\bar\phi\phi^2\right]+I_{\bar\phi}=0
\label{gapphib}
\end{equation}
with,
\begin{equation}
I_{\bar\phi}=\frac{\partial\Omega_{\bar q q}}{\partial \bar\phi}=-6N_fT\int\frac{d\zbf p}{(2\pi)^3}
\left[\frac {e^{-2\beta\omega_-}}{1+3\phi e^{-\beta\omega_-}+3\bar\phi e^{-2\beta\omega_-} + e^{-3\beta\omega_-}}
+\frac {e^{-\beta\omega_+}}{1+3\phi e^{-\beta\omega_+}+3\bar\phi e^{-2\beta\omega_+} + e^{-3\beta\omega_+}}
\right],
\end{equation}
By solving Eqs.(\ref{gapsigma}),(\ref{gapphi}) and (\ref{gapphib}) self consistently one can get the values of constituent quark mass,
Polyakov loop variable and the conjugate Polyakov loop variable as a function of temperature. 

\section{Scattering amplitudes}
\label{scatterings}
\begin{figure}[tbh]
\subfigure{
\includegraphics[width=3.5cm]{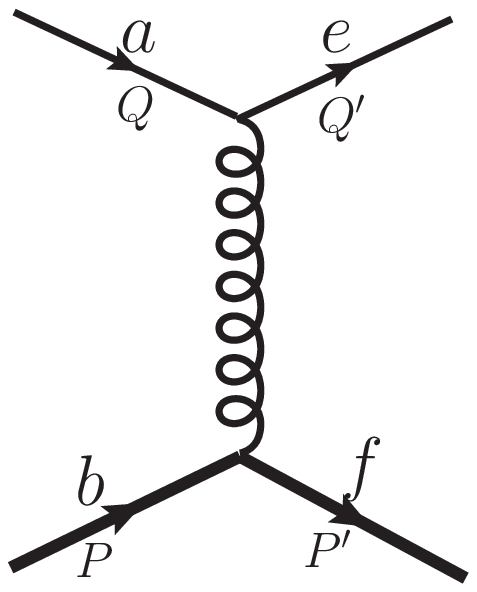}}
\subfigure{
\hspace{-0mm}\includegraphics[width=3.5cm]{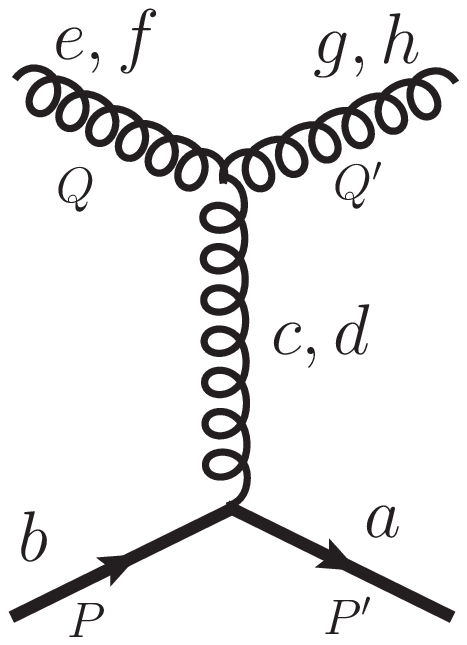}} 
\caption{Coulomb scattering (left) of HQ (bold solid line) and light quark/antiquark (thin solid line). t-channel Compton scattering (right) }
\label{st}
\end{figure}

There are two types of scatterings that contribute to the drag and the diffusion coefficients namely Coulomb scattering i.e., scattering off of HQ from light quark and Compton scattering i.e., scattering off of gluon from HQ ~\cite{Svetitsky:1987gq}. The dominant contribution for these scatterings arise from the gluon exchange in the t-channel which is infrared divergent \cite{Rapp:2009my,Moore:2004tg}. This is regularised by introducing the Debye screening~\cite{Moore:2004tg, Svetitsky:1987gq} which we have evaluated in the HTL limit in the background of Polyakov loop. In s and u channel, however, there is no such infrared divergences. Note that in the matrix model, $m_D$, in Eq.(\ref{mdb}) is color dependent so the propagator is also color dependent. For $N_f$ flavor of light quark,  the spin averaged matrix element squared for Coulomb scattering  as shown on the left side of Fig.(\ref{st}), can be written as
\begin{equation}
|\mathcal{M}_C|^2=\frac{16 N_fg^4}{ 8N}\mathcal{P}^{cd}_{ae}\mathcal{P}^{ml}_{bf}\mathcal{P}^{c'd'}_{ea}\mathcal{P}^{m'l'}_{fb}\frac{((s-m^2-M^2)^2+(u-m^2-M^2)^2+2(M^2+m^2)t)}{(t+(m_D^2)_{mlcd})(t+(m_D^2)_{m'l'c'd'})}.
\label{coulomb21}
\end{equation}
where $a,b(e,f)$ are color indices of initial (light,heavy) and final (light,heavy) quarks. 
For calculational simplifications, one can take the color averaged Debye mass as defined in Eq.(\ref{dbmass1}) so that $(m_D^2)_{mlcd} \approx \bar{m}_D^2 \mathcal{P}_{mlcd}$. In this case, we get
\begin{eqnarray}
\mathcal{P}^{cd}_{ae}\mathcal{P}^{ml}_{bf}\frac{1}{(t+(m_D^2)_{mlcd})}=\frac{1}{t+\bar{m}_{D}^2\mathcal{P}^{fb}_{ae}}-\frac{1}{N}\bigg(\frac{2}{t+\bar{m}_D^2}-\frac{1}{N}\frac{1}{t+\bar{m}_{D}^2}\bigg)\delta_{ae}\delta_{fb}
\end{eqnarray}
One can further simplify the expression in Eq.(\ref{coulomb21}) by taking the leading order contribution in $N$. With this assumption, Eq. (\ref{coulomb21}) reduces to
\begin{equation}
|\mathcal{M}_C|_{abef}^2=\frac{8g^4}{2 N}\delta_a^f\delta_e^b
\frac{((s-m^2-M^2)^2+(u-m^2-M^2)^2+2(M^2+{m}^2)t)}{(t+(\bar{m}_D^2)^2)^2}
\label{coulomb2}
\end{equation}
where $M$ is HQ mass. For the $q^{a} Q^{b}\rightarrow q^{e}Q^{f}$ scattering, the product of distribution function and matrix element squared that appears in Eq.(\ref{transport111}) can be simplified by summing over color of initial and final light/heavy quarks. Note that for light quarks, the colors appearing in Eq.(\ref{coulomb2})  has to be summed with the distribution function and can be written as
\begin{equation}
\delta_a^f\delta_e^b f(q)_{e}(1-f(q')_{f})=N^2f(q)_{q}(1- f(q')_{q})
\label{proj}
\end{equation}
where $f(q)_{q}$ is the average distribution function of quark as defined in Eq.(\ref{avgfunq}). Similarly for the $t$ channel Compton scattering shown on the right side of Fig.(\ref{st}), one can write
\begin{equation}
|\mathcal{M}_t|^{2}=\frac{g^4}{4(N^2-1)}\mathcal{P}^{ml}_{ba}\mathcal{P}^{l'm'}_{ab}f^{cd,ef,gh}f^{d'c',fe,hg}\bigg(\frac{16(s-M^2)(M^2-u)}{(t+(m_D^2)_{mlcd})(t+(m_D^2)_{m'l'c'd'})}\bigg).
\end{equation}
and can be simplified in a similar way as done for Coulomb scattering. Here $ef,b(gh,a)$ are the color indices for initial (final) gluon and quark. 
\begin{figure}[tbh]
\subfigure{
\includegraphics[width=5.5cm]{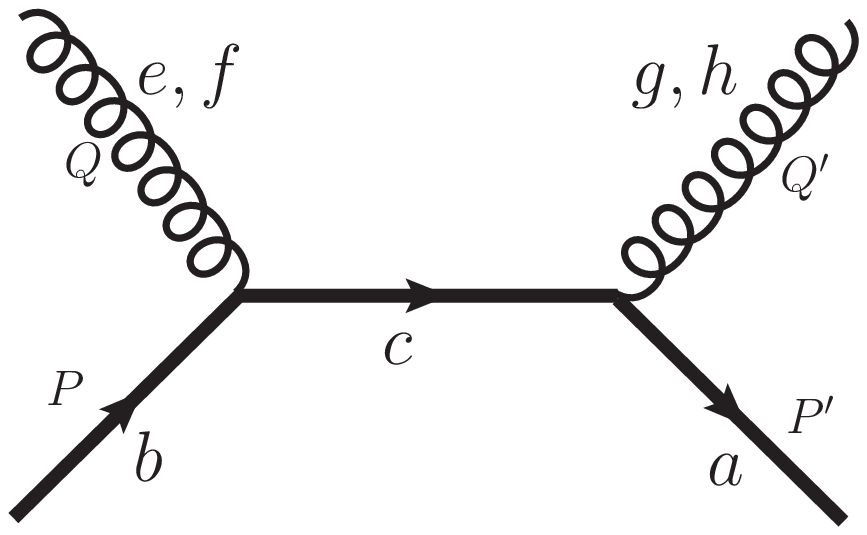}}
\subfigure{
\hspace{-0mm}\includegraphics[width=5.5cm]{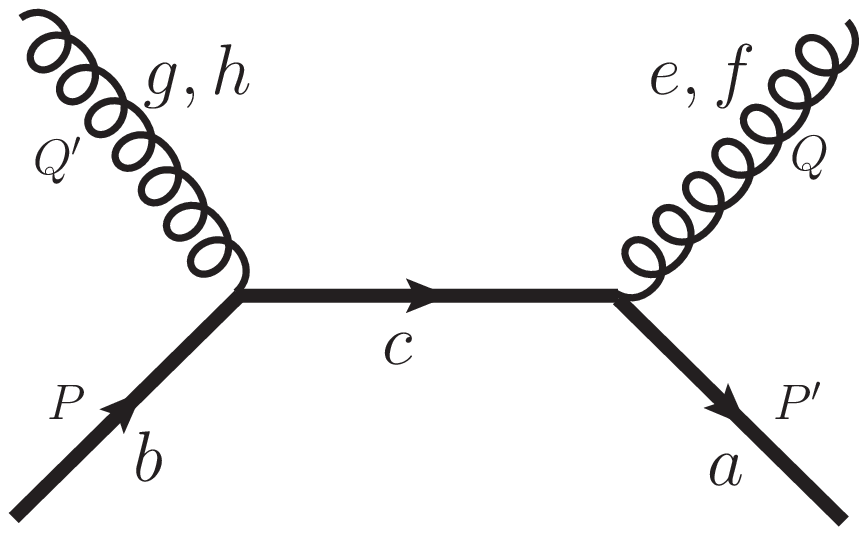}} 
\caption{s-channel Compton scattering (left). u-channel Compton scattering (right)}
\label{su}
\end{figure}
The scattering amplitude of u channel Compton scattering shown on the right side of Eq.(\ref{su}) can be written as
\begin{equation}
|\mathcal{M}_u|^{2}=\frac{8g^4}{8(N^2-1)}\mathcal{P}^{gh}_{bc}\mathcal{P}^{gh}_{bc'}\mathcal{P}^{ef}_{ca}\mathcal{P}^{ef}_{c'a}\bigg(\frac{M^4-us+M^2(3u+s)}{(u-M^2)^2}\bigg).
\label{uch1}
\end{equation}
Note here that the propagator has no color dependent term.  Matrix element squared for s channel Compton scattering as shown on the left side of Fig.(\ref{su}) is
\begin{equation}
|\mathcal{M}_s|^{2}=\frac{8g^4}{8 (N^2-1)}\mathcal{P}^{ef}_{bc}\mathcal{P}^{ef}_{bc'}\mathcal{P}^{gh}_{ca}\mathcal{P}^{gh}_{c'a}\bigg(\frac{M^4-us+M^2(u+3s)}{(s-M^2)^2}\bigg).
\label{ms1}
\end{equation}
There are interferences between different scatterings contributing to $g^{ef}Q^{b}\rightarrow g^{gh}Q^{a}$ that can be written as
\begin{equation}
\mathcal{M}_{s}{\mathcal{M}_u}^\dagger=\mathcal{M}_{u}{\mathcal{M}_s}^\dagger=\frac{g^4}{8(N^2-1)}\mathcal{P}^{ef}_{bc}\mathcal{P}^{gh}_{ca}\mathcal{P}^{gh}_{bc'}\mathcal{P}^{ef}_{c'a} \bigg(\frac{32M^4-8M^2t}{(s-M^2)(u-M^2)}\bigg).
\label{suit1}
\end{equation}
\begin{equation}
\mathcal{M}_{s}{\mathcal{M}_t}^\dagger=\mathcal{M}_{s}^\dagger{\mathcal{M}_t}=\frac{g^4}{4\sqrt{2}(N^2-1)}\mathcal{P}^{ef}_{bc}\mathcal{P}^{gh}_{ca}\mathcal{P}^{lm}_{ab}(if^{dc,fe,hg}) \bigg(\frac{-8(M^4-2M^2s+us)}{(s-M^2){(t+(m_D^2)_{mlcd})}}\bigg).
\end{equation}
\begin{equation}
\mathcal{M}_{u}{\mathcal{M}_t}^\dagger=\mathcal{M}_{u}^\dagger{\mathcal{M}_t}=\frac{g^4}{4\sqrt{2}(N^2-1)}\mathcal{P}^{gh}_{bc}\mathcal{P}^{ef}_{ca}\mathcal{P}^{lm}_{ab}(if^{dc,fe,hg}) \bigg(\frac{8(4M^4-M^2t)}{(u-M^2){((t+(m_D^2)_{mlcd}))}}\bigg).
\end{equation}
Total matrix element squared that contribute to Compton scattering i.e., $gQ\rightarrow gQ$ is $|\mathcal{M}_{Cm}|_{abefgh}^2=|\mathcal{M}_s|^2+ |\mathcal{M}_u|^2+|\mathcal{M}_t|^2+\mathcal{M}_u \mathcal{M}_s^{\dagger}+\mathcal{M}_s\mathcal{M}_{u}^{\dagger}+\mathcal{M}_t\mathcal{M}_s^{\dagger}+\mathcal{M}_s\mathcal{M}_{t}^{\dagger}+\mathcal{M}_u\mathcal{M}_t^{\dagger}+\mathcal{M}_t\mathcal{M}_{u}^{\dagger}$. These matrix elements are used in Eq.(\ref{transport111}) to estimate the drag and the diffusion coefficient.
\vspace{2cm}
\vspace{2cm}

\def\bjschaefer{B. J. Schaefer, J. M. Pawlowski and J. Wambach, {\PRD{76}{074023}{2007}}.}
\def\guptatiwari{U.S. Gupta, V.K. Tiwari,{\PRD{85}{014010}{2012}}.}
\def\bielich{B.W. Mintz, R.Stiele, R.O. Ramos, J.S. Bielich,{\PRD{87}{036004}{2013}}}
\def\ranjita{H. Mishra, R.K. Mohapatra,{\PRD{95}{094014}{2017}}.}
\def\buballa{S. Carignano, M. Buballa, W.Elkamhawy,{\PRD{94}{034023}{2016}}}

\end{document}